\documentclass{article}
\pdfoutput=1

\makeatletter
\newif\if@restonecol
\makeatother

\usepackage{graphicx}
\usepackage{latexsym}

\usepackage{amssymb}
\usepackage{xspace}
\usepackage[ruled,vlined]{algorithm2e}
\usepackage{algorithmic}
\usepackage{verbatim}

\usepackage{times}
\usepackage[latin1]{inputenc}
\usepackage{color}
\usepackage{epsfig}
\usepackage{subfigure}
\usepackage{multirow}

\newcommand{\instance}{\mbox{${\it inst\/}$}}

\def\blackbox{{\rule{0.5mm}{0.5mm}}}
\def\qed{\hspace*{\fill}\blackbox}
\newcommand{\condbox}{\qed}

\newtheorem{lemmc}{Lemma}
\newtheorem{proposition}{Proposition}
\newtheorem{example}{Example}
\newtheorem{defn}{Definition}

\newcommand{\crunchbegin}{}
\newcommand{\crunchend}{}   

\def\blackbox{{\rule{1.5mm}{1.5mm}}}

\newcommand{\tld}{$~$}

\newcommand{\removesmallspace}{\vspace*{-0.18cm}}

\newcommand{\schema}[1]{{\textbf{#1}}}

\newcommand{\pair}[2]{\left\langle #1, #2 \right\rangle}

\newcommand{\scenario}[1]{{\cal #1}}

\newcommand{\vars}[1]{{\overline{#1}}}

\newcommand{\SOL}{\mathsf{Sol}}

\newcommand\M{\scenario M}

\usepackage{amssymb}

\begin{document}

\title{Semantic Query Reformulation in Social PDMS}

\author{Angela Bonifati\\
\small
LIFL, University of Lille 1\\
\small
Cit\'e Scientifique, Lille (France)\\
\and 
Gianvito Summa\\
\small
University of Basilicata\\
\small
Via dell'Ateneo Lucano, Potenza (Italy)\\ 
\and 
Esther Pacitti\\
\small
LIRMM, University of Montpellier II\\
\small
Rue Ada, Montpellier (France) 
\and 
Fady Draidi\\
\small
LIRMM, University of Montpellier II\\
\small
Rue Ada, Montpellier (France)
}

\maketitle

\begin{abstract}
We consider social peer-to-peer data management systems (PDMS),
where each peer maintains both semantic mappings between its schema and some acquaintances, and social links with peer friends.
In this context, reformulating a query from a peer's schema into other peer's schemas is a hard problem, as it may generate as many rewritings as the set of mappings
from that peer to the outside and transitively on, by eventually traversing the entire network.
However, not all the obtained rewritings are relevant to a given query.
In this paper, we address this problem by inspecting semantic mappings and social links to find only relevant rewritings.
We propose a new notion of `relevance' of a query with respect to a mapping,
and, based on this notion, a new semantic query reformulation approach for
social PDMS, which achieves great accuracy and flexibility.
To find rapidly the most interesting mappings, we combine several techniques:
\emph{(i)} social links are expressed as FOAF (Friend of a Friend) links to characterize peer's friendship
and compact mapping summaries are used to obtain mapping descriptions; \emph{(ii)} local semantic views are 
special views that contain information about external mappings; and \emph{(iii)} gossiping techniques improve the search of relevant mappings.
Our experimental evaluation, based on a prototype on top of PeerSim
and a simulated network demonstrate that our solution
yields greater recall, compared to traditional query translation approaches proposed in the literature. 
\end{abstract}

\section{Introduction}
In the last decade, we have witnessed a dramatic shift in the scale of 
distributed and heterogeneous databases: they have become larger, more 
dispersed and semantically interconnected networks of peers, exhibiting
varied schemas and instances.  
This shift in scale has forced us to revisit the assumptions underlying 
distributed databases and consider peer-to-peer (P2P) systems. 
A P2P data management system (PDMS)~\cite{Halevy05} is an ad-hoc collection of 
independent peers that have formed a network in order to map and share 
their data. For example, consider an  
online scientific community\footnote{This example 
has been inspired by a web-based online trusted physician network, https://www.ozmosis.com/home, 
`where good doctors go to become great doctors'.}, 
that uses an underlying P2P infrastructure 
for data sharing. In particular, each peer embodies a medical doctor or a physician, 
who enters the community to share her clinical cases (yet hiding sensitive patient record data) with a subset of her colleagues, and get knowledgeable opinions from them.

Peers in such example typically have heterogeneous schemas, with 
no mediated or centralized schema. Still, to process a query over the PDMS, the data needs to be translated from one 
peer's schema to another peer's schema. 
To address this problem, PDMS maintains a set of mappings or correspondences between a
peer schema and a sufficiently small number of other peer schemas, called {\em acquaintances}.
The mappings between the local schema and the acquaintance 
schema can be manually provided, or, alternatively, computed via an external schema matching tool. 


Each doctor likes to exchange specific data about treatments and patients
with the peers she trusts, and/or she is friend with. Additionally, she may not find the information within her set of acquaintances, and may need to look for colleagues she has never met before.

In order to cope with data heterogeneity in PDMS, queries are formulated against 
a local peer schema, and translated against each 
schema of the peer acquaintances, and transitively on. This problem, called {\em query reformulation}, 
has been addressed in the literature by schema mappings tools~\cite{Popa2002, Yu2004, Bonifati2010}, and proved to be effective in PDMS~\cite{Ives2004}. 
However, a fundamental limitation of the above tools is the fact that 
query translation is essentially enacted on every peer by tracking all the mappings, whereas in a realistic scenario, only semantically relevant mappings must be exploited.
E.g. in our online community, each doctor would like to 
exchange specific data about treatments and patients only 
with the peers that provide relevant information (members of the same lab or former university mates), rather than with every peer. 
Similarly, she may be willing to know who else, among the doctors in her community, or among her friend doctors, has worked on similar cases.

As the above example (typical of professional social networking) suggests, social relationships (or friendships) between community members are also crucial to locate relevant information. 
Similarly, in order to identify relevant mappings, we exploit friendship links between peers, in addition to acquaintances,
in what we call social PDMS. As in social networks, by establishing a friendship link, a peer $p_i$ can become friend with a peer $p_j$ and share peer information.
In our case, the peer information we are interested in 
is local semantic mappings. They express meaningful semantic relationships between elements in heterogeneous schemas on different peers. Local mappings can be copious. For such a reason and for efficient storage and retrieval at peers, 
local mappings may need to be summarized as Bloom filters. 
In addition, to capture peer friendship, we adopt the Friend of a Friend (FOAF) vocabularies~\cite{Berners2009}. FOAF provides an open, detailed description of users profiles and their relationships using RDF syntax. 
We have adapted the FOAF files to PDMS and extended the FOAF syntax to also point to the mapping summaries
of a peer's friends.

In this paper, we tackle the problem of query reformulation for
conjunctive queries in social PDMS. 
Based on a new notion of {\em relevance} of a query with respect to a mapping, we propose a semantic query reformulation approach, using both semantic mappings and friendship links, thus biasing the query translation
only towards {\em relevant} peers.



To precisely define the notion of relevance of a query with respect to a mapping, we propose a novel metric called AF-IMF measure, which takes into account the semantic proximity between the query and the local and external 
mappings. 
However, the above metric would need to be computed distributively, and to do so, would have to  
contact every peer in the network.
To address this difficulty, we do store on each peer a local semantic view, that offers a synthetic description of the mapping components of 
external peers. To further improve the search of relevant mappings, we combine the above local semantic views with 
gossiping techniques~\cite{KermarrecS07}.
These techniques refer to the probabilistic exchange of 
mappings between two peers, thus leading to the endless process of 
making two random peers communicate among each other. 
We adapt gossiping to our context by periodically refreshing the local semantic view on each peer, 
based on gossiped atoms; by means of such semantic views, promising semantic paths can be undertaken in the 
network, such that, for a given query, the most relevant mappings can 
be located and/or the most relevant peer friends can be reached. 


\noindent {\bf Contributions.}
We make the following main contributions:

\emph{(i)} We propose a novel notion of relevance of query with respect to a mapping, along with that of a relevant rewriting; we characterize each mapping in the entire collection of mappings present in the network with a new metric, the AF-IMF measure, which precisely 
identifies the most interesting mappings, towards which query translation should be directed.

\emph{(ii)} We propose query reformulation algorithms that, given 
an input query $Q$, and a set of mappings between peers schemas, do the following: translate the query into $Q^t$ only against the relevant mappings 
by adopting our new evaluation metric;
exploit friendship links among peers to possible enlarge the 
set of mappings and bias the search towards interesting peers;
exploit semantic gossiping to discover new relevant 
mappings and friends, thus increasing the number of query rewritings.
To the best of our knowledge, these algorithms advance the state of art of query reformulation in PDMS (more 
details in Section~\ref{sec:rel}).

\emph{(iii)} We provide an extensive experimental evaluation
by running our algorithms on a simulated network built on top of PeerSim, which demonstrates that our solution
yields greater recall, compared to traditional
query translation approaches.

The paper is organized as follows. Section~\ref{sec:prelim} presents the background and the problem definition. Section~\ref{sec:frame} introduces our 
framework, while Section~\ref{sec:algos} and 
Section~\ref{sec:expe} describe our algorithms and the experimental assessment that has been conducted. Finally, Section~\ref{sec:rel}
discusses the related work and Section~\ref{sec:concl} concludes the paper.

\section{Problem Definition}
\label{sec:prelim}

In this section, we first present the background of schema mappings and P2P networks, and then we detail the problem statement.
     
\subsection{Schema Mapping Model}
Data exchange systems~\cite{Fagin2009} rely on dependencies to specify mappings.
Given two schemas, $\schema{S}$ and $\schema{T}$, a \emph{source-to-target tuple--generating dependency} (also called a s-t tgd or, equivalently, a tgd) is a first-order formula of the form $\forall \vars{x} (\phi(\overline{x}) \rightarrow \exists \vars{y} (\psi(\vars{x}, \vars{y}))$, where $\vars{x}$ and $\vars{y}$ are vectors of variables, $\vars{x}$ are universally quantified variables and 
$\vars{y}$ are existentially quantified variables. The body $\phi$ is a conjunctive query (CQ) over $\schema{S}$ and the head $\psi$ is a CQ over $\schema{T}$.

\begin{example}
Consider Figure~\ref{fig:schemaMappingExample}
in which two schemas describing two scientists' local data are depicted.
A set of correspondences $v_1, v_2$ and $v_3$ connects elements in the two schemas. 
\begin{figure}[t!]
\begin{center}
\includegraphics[width=7cm]{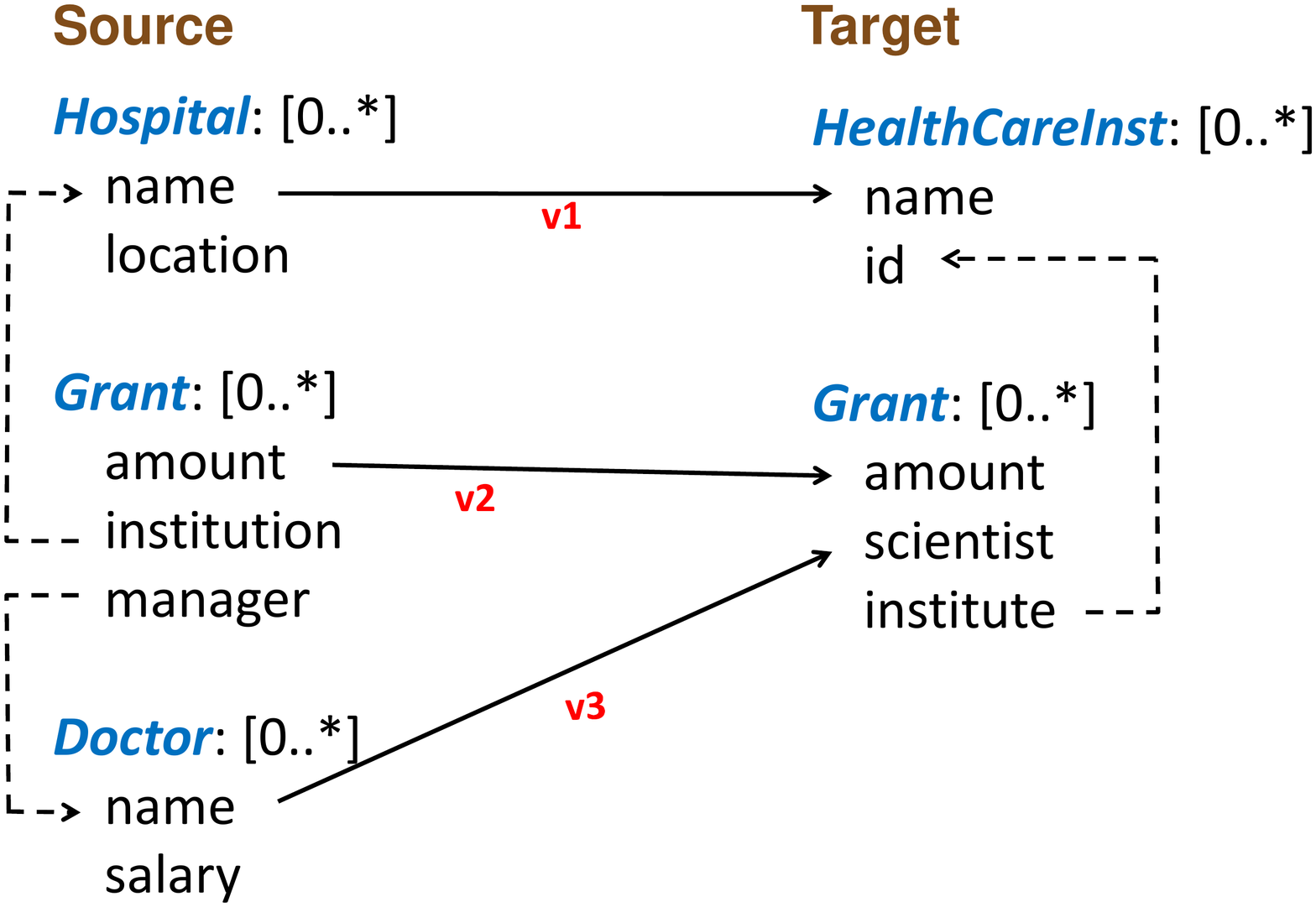}
\caption{A Schema Mapping Example}
\label{fig:schemaMappingExample}
\end{center}
\end{figure}

We report below two examples of s-t tgds for the two schemas above:

\removesmallspace
\[
\begin{array}{l}
\textsc{Source-to-Target Tgds} \\
m_1.\tld \forall n, l\!: \emph{Hospital}(n, l) \rightarrow \exists I\!\!: \emph{HealthCareInstitution}(n, I) \\
m_2.\tld \forall n, s, a, pi, l\!: \emph{Doctor}(n, s) \wedge \emph{Grant}(a, pi, n) \wedge \emph{Hospital}(pi, l) \\
\quad \quad   \rightarrow \exists I\!\!: \emph{HealthCareInstitution}(pi, I) \wedge \emph{Grant}(a, n, I)\\
\end{array}
\removesmallspace
\]
\end{example}

A \emph{schema mapping} is a triple $\scenario{M} = (\schema{S}, \schema{T}, \mu_{st})$ ($\M_{st}$, in short), where $\schema{S}$ 
is a source schema, $\schema{T}$ is a target schema, $\mu_{st}$ is a set of source-to-target tgds. If $I$ is 
an instance of $\schema{S}$ and $J$ is an instance of $\schema{T}$, then the pair $\pair{I}{J}$ is an instance 
of $\pair{\schema{S}}{\schema{T}}$.
A target instance $J$ is a \emph{solution} of $\scenario{M}$ and a source instance $I$  
(denoted $J \in \SOL(\scenario M, I)$) iff $\langle I, J \rangle \models \mu_{st}$, 
i.e., 
$I$ and $J$ together satisfy the dependencies.



We distinguish between specific forms of s-t tgds,
which are GAV (global-as-view) and LAV (local-as-view). A GAV tgd is a formula 
$\forall \vars{x} (\phi(\overline{x}) \rightarrow {\cal A}(\vars{x}))$, where the head is a single atom 
${\cal A}(\vars{x})$. Similarly, a LAV tgd is a formula 
$\forall \vars{x} ({\cal A}(\overline{x}) \rightarrow \exists \vars{y} (\psi(\vars{x}, \vars{y}))$, where the
 body is a single atom ${\cal A}(\vars{x})$. GAV tgds are special cases of more general tgds, 
 called GLAV, that contains conjunctions of atoms and existential variables in the head. 
Here and henceforth, we focus on GLAV mappings, thus expressed by means of GLAV s-t tgds in $\mu_{st}$, 
as the others represent more restrictive cases. 
We denote such GLAV mappings with $\scenario{M}$.

\vspace{-2mm}
\begin{example} [cont'd]
Continuing with the above example, a schema mapping is a triple 
$\scenario{M} = (\schema{S}, \schema{T}, \mu_{st})$, where $\schema{S}$ is the source schema, $\schema{T}$ 
is the target schema and 
$\mu_{st} = \{m_1, m_2\}$. Moreover, $\M$ is a GLAV mapping.  
\end{example}

We assume as customary that mappings among schemas are either provided by the users or by using external schema mapping tools.

We build on prior work~\cite{Yu2004, Bonifati2010} to define the semantics of query translation. Precisely, we denote with $\instance(\schema{S})$ (\instance(\schema{T}))
the set of instances $I$ (instances $J$, respectively).
 
\begin{defn}[Semantics of query translation]
\label{def-sem}
{\em Suppose $Q_i$ is a query posed against $\schema{S}$, and 
$Q_j$ is a query posed against $\schema{T}$, $j \neq i$. 
Let $Q_i^t$ denote a translation of $Q_i$ against $\schema{T}$ and 
$Q_j^t$ denote a translation of $Q_j$ against $\schema{S}$.
Then, $Q_j^t$ is {\em correct} provided $\forall D_s \in \instance(\schema{S}): \;
Q_j^t(D_s) = Q_j(\M(D_s))$.
The translation $Q_i^t$ is {\em correct} provided $\forall D_t \in \instance(\schema{T}): \;
Q_i^t(D_t) = \bigcap_{D_s^k: \M(D_s^k) = D_t}  \, Q_i(D_s^k))$. 
}
\end{defn}


In other words, the translation $Q_j^t$ is correct provided 
$Q_j$ applied to the transformed instance $\M(D_s)$ and $Q_j^t$ applied to $D_s$ both yield
the same results, for all $D_s \in \instance(\schema{S})$. Note that in this case, the
direction of translation is {\sl against} that of the mapping $\M$.
As in~\cite{Bonifati2010}, we henceforth call it {\em backward query translation}. 
This direction of translation is similar to view expansion, with $\M$
being the view definition.
Translating a query $Q_i$ posed against $\schema{S}$ to the schema
$\schema{T}$ of peer $p_j$ is {\sl aligned} with the direction of the
mapping $\M$, and represents the {\em forward query translation}~\cite{Bonifati2010}.
The forward direction is more tricky, as the mapping $\M$ may not be invertible.
In fact, there are two alternative strategies to make sense of this direction of 
translation: \emph{(i)} obtaining the reverse schema mapping $\M^{-1}$~\cite{Fagin2007b,Arenas2008a}, 
such that the query rewriting semantics is the same as the backward direction;
this strategy applies to the case in which one would like to recover the 
exchanged data, i.e. to find the source instance $I$ from which the target instance $J$ has been derived;
\emph{(ii)} focusing on the computation of a rewriting of a conjunctive query 
$Q_1$ over the source schema, assuming that a source instance $I$ ($D_s$) is already available and adopting the 
semantics based on certain answers of all possible pre-images $D_s^k$; in such a case, it is possible to reuse the work done in 
the area of query answering using views for data integration~\cite{Levy1995, Lenzerini2002}.

Moreover, observe that in our setting we focus on query answering rather than on data exchange and on materializing a target instance. In fact, by following the semantics given 
in ~\cite{Fagin2005a, Bonifati2010}, we adopt the second strategy \emph{(ii)}, that lets us translate the query rather than the data and  
lets us realize query rewriting along the mappings. This strategy is more natural in a P2P setting in which we do not need to
  reverse the mappings, and lets us avoid bringing the exchanged data back to the peers.



\subsection{Network Model} 
We assume a heterogeneous network of peers $p_1, \dots, p_n$, each peer having a 
distinct relational~\footnote{The extension 
to nested relational schemas is beyond the scope of our paper and will be addressed as future work.} schema $S_1,$ $\dots, S_n$. 
Let $\M_{ij}$ be a generic GLAV
mapping between a pair of schemas $S_i,\ S_j$, from peer $p_i$ to peer $p_j$. 
We assume that 
each peer has only one local schema, which may contain key/foreign key constraints, along with data defined 
according to the schema itself. However, for simplicity we ignore the above constraints in the query translation process.

Given a mapping $\M$ from a peer $p_i$ to a peer $p_j$, which we 
denote with $\M_{ij}$, $\M_{ij}$ is also called an {\em outward mapping} for 
$p_i$. The peer $p_j$ is also called the {\em target peer} for this mapping. 
By opposite, a mapping  $\M_{ji}$ from peer $p_j$ to $p_i$ is called an 
{\em inward mapping} for $p_i$ ({\em outward} for $p_j$, resp.). Similarly, $p_i$ is the target peer 
in such a case.

We do not assume a symmetric distribution of the mappings, i.e. with a mapping $\M_{ij}$, we expect that 
either $p_i$ (resp. $p_j$) stores the mapping or both of them.
We have designed ad-hoc data structures to store mappings on each peer.
Details on such data structures will be provided in Section~\ref{sec:datastruct}. 
As customary, the network has a
dynamic behavior, meaning that any peer $p_i$ can join or leave the network 
arbitrarily.
 

\subsection{Problem Statement}
Without loss of generality, we consider conjunctive queries (CQs), that are expressed as
conjunctions of atoms $a_1, \cdots, a_n$, and mappings $\M$ composed by mapping rules having one or more atoms in the body $\phi$ 
 (in the head $\psi$, respectively).


Given an input query $Q_i$ formulated at a peer in the network against an arbitrary schema $S_i$,  and a direct outward mapping $\M_{ij}= S_i \rightarrow S_j$ (from $S_i$ to $S_j$) and a direct inward mapping $\M_{ki} = S_k \rightarrow S_i$ (from $S_k$ to $S_i$) and, in addition, transitively 
from $S_j$ (resp. $S_k$) to any other reachable schema $S_l$ (resp. $S_m$) for which it exists, without loss of generality, at least an 
outward mapping $\M_{jl}$ (resp. an inward mapping $\M_{mk}$) and so on, continuing from $S_l$ and $S_m$ to any other reachable schema through inward or outward mappings.

Then, the problem can be stated as: 
\begin{itemize}
\item finding the relevant rewritings of $Q_i$ along and against the 
direction of the mappings $\M_{ij}$ ($\M_{ki}$, resp.) and  $\M_{jl}$ ($\M_{mk}$, resp.)
and so on, by following the mappings which connect the schemas.
All the relevant rewritings have to be computed by avoiding useless mapping paths from $S_i$ to $S_j$, from $S_k$ to $S_i$, from $S_j$ to $S_l$, from $S_m$ to $S_k$ and so on, from any reached schema to any reachable schema connected by mappings.
\end{itemize}


Notice that the propagation of the input query $Q_i$ to all peers in the network leads to collect
as many rewritings as possible for that query. In fact, the input query $Q_i$ can be certainly evaluated on 
the originating peer that hosts the schema $S_i$ (upon which the query itself has been formulated) but may not be pertinent for all the schemas of other peers, unless relevant rewritings can be located. 
Moreover, the chosen strategy by which the results of the rewritten queries are conveyed towards the originating peer is 
a simple one, i.e. the results are unioned and possible duplicates are discarded.
Alternatively, mapping composition could have been used here, but it falls beyond the scope of this paper. 


The rewritings of $Q_i$ follow the semantics given in Definition 1, whose
correctness is proved in~\cite{Bonifati2010}.
In this paper, we propose a query rewriting strategy different from the ones used in previous work~\cite{Ives2004,Bonifati2010} in which all possible translations are pursued, since we only exploit relevant translations. To this purpose, Section~\ref{sec:frame} introduces the notion of relevance of a query with respect to a mapping, and that of a relevant rewriting.


\section{A Framework for query reformulation}~\label{sec:frame}
\vspace{-2mm}

In this section, we develop a novel framework for
semantic query reformulation in social PDMS.
This framework relies on several contributions: a precise definition of relevance of a query wrt. a mapping; a new metric (AF-IMF) for computing such relevance and its supporting data structures; and a distributed method for computing AF-IMF in a P2P network.

\subsection{Relevance of a query wrt. a mapping}


To define such relevance, we consider a schema mapping scenario
$\scenario{M} = (\schema{S}, \schema{T}, \mu_{st})$, where $\schema{S}$ is a source schema, $\schema{T}$ is 
a target schema, $\mu_{st}$ is a set of source-to-target tgds that express the GLAV mapping.

Let $m \in \mu_{st}$ be a s-t tgd~\footnote{Notice that here and henceforth we use \emph{mapping rule} and \emph{s-t tgd} as synonyms.} of the form 
$\forall \vars{x} (\phi(\overline{x}) \rightarrow \exists \vars{y} (\psi(\vars{x}, \vars{y}))$, 
with $\phi$ and $\psi$ as CQ queries, containing the atoms
$a_1(X_1),$ $\cdots, a_n(X_n)$, with each $X_i$ 
being an ordered set of parameters ($x_1$,$x_2$, $\cdots$, $x_i$), and each parameter
being a variable $\$x_1$.

Let $Q$ be a CQ containing the atoms $a_1(X_1),$ $\cdots, a_n(X_n)$, with each $X_i$ 
being an ordered set of parameters ($x_1$,$x_2$, $\cdots$, $x_i$), and each parameter
being a constant value $x_1$ or a variable $\$x_1$.

\noindent 
A query atom $a_i(X_i)$ in $\phi$ ($\psi$, resp.) is unifiable with a
query atom $a_j(X_j)$ in $Q$ if a unifying substitution of variables and constant symbols exists. 
More precisely, a unification occurs if:

\begin{itemize}
\item \emph{(i)} label($a_i$) = label($a_j$), i.e. both atoms have the same name; 
\item \emph{(ii)} $\forall x_i \in X_i$, $\$x_i$
matches the variable symbol $\$x_j \wedge (i=j)$ or $\$x_i$ matches the constant symbol $x_j \wedge (i=j)$.   
\end{itemize}

In other words, each query atom $a_i$($X_i$) must match an atom in the body (head, resp.) of a tgd with both 
its label and its set ordered of parameters ($x_1$, $\cdots$, $x_m$). Such a match follows the 
rules for atoms unification in Datalog (i.e. constant and 
variable unification). 

\begin{example}
Consider again Figure~\ref{fig:schemaMappingExample} and the mapping rules
$m_1$ and $m_2$ specified in Section 2. 


If we consider as a query $Q = \emph{Hospital}(\$x, 'San Francisco')$,
this query being posed against the source schema of Figure~\ref{fig:schemaMappingExample}, returns 
the names of all hospitals in San Francisco.
$Q$ consists of only one atom (Hospital) which has two parameters, a variable (i.e. $\$x$) plus a constant 
value (i.e. "San Francisco").
\end{example}


We can now define the relevance of a query with respect to a mapping rule, as follows.

\begin{defn}[Relevance Forward]
Given a schema mapping $\M_{ij}$ that maps elements of the 
schema $S_i$ into elements of $S_j$ and let $m$ be a mapping rule in $\mu_{ij}$, let ${\cal A}_{i}$ be the set of atoms of $m$ in the body, 
a query $Q$ posed against $S_i$ along the direction of the mapping rule is relevant to $m$ if 
$\forall a_q$ 
of $Q$,  $a_q \in {\cal A}_{i}$, i.e. each atom of $Q$ is unifiable with an atom of ${\cal A}_{i}$.
\end{defn}

\begin{defn}[Relevance Backward]
Given a schema mapping $\M_{ij}$ that maps elements of the 
schema $S_i$ into elements of $S_j$ and let $m$ be a mapping rule in $\mu_{ij}$, 
let ${\cal A}_{j} = \{ a_j(X_j)\}$ be the set of atoms of $m$ in the head, 
such that $a_j(X_j) \in {\cal A}_{j}$ if it only contains universally 
quantified variables, 
a query $Q$ posed against $S_j$ against the direction of the mapping rule is relevant to 
$m$ if $\forall a_q$ of $Q$,  $a_q \in {\cal A}_{j}$, i.e. each atom of $Q$ is unifiable with 
an atom of ${\cal A}_{j}$.
\end{defn}

Consequently, we can now define the relevance of a query wrt. the whole mapping, as follows.

\begin{defn}[Mapping Relevance]
Let $\scenario{M} = (\schema{S}, \schema{T}, \mu_{st})$ be a mapping,  where $\schema{S}$ 
is a source schema, $\schema{T}$ is a target schema and $\mu_{st}$ be the set of s-t tgds and let $m \in \mu_{st}$ be a mapping rule of such mapping.
A query $Q$ posed against the mapping $\M$ is relevant if there exists at least one mapping rule $m \in \mu_{st}$ so that $m$ is forward or backward relevant for $Q$. 
\end{defn}

\begin{example}[cont'd]
Continuing with the example above, shown in Figure~\ref{fig:schemaMappingExample},
it is easy to check that the query $Q$ above is forward relevant to both $m_1$ and $m_2$, according to the above 
definition. 
If we consider a query $Q' = \emph{Grant}(\$x, \$y, \$z)$ and a query 
$Q'' = \emph{HealthCareInstitution}(\$y, \$z)$,
neither $Q'$ nor $Q''$ are backward relevant to either mapping rule.
\end{example}


Here and henceforth, we will use the term \emph{relevance} to denote mapping relevance, unless otherwise specified. 
It follows that, if a query $Q$ is relevant to a mapping $\M_{ij}$, 
its translation $Q_i^t$ is also relevant to that mapping.

\vspace{-2mm}
\begin{proposition}
If a query $Q$ formulated against $S_i$ is relevant to a mapping $\M_{ij}$, 
its translation $Q^t$ formulated against $S_j$ is also relevant to $\M_{ij}$, 
and viceversa.
\end{proposition}

The proof of the above proposition is straightforward and is omitted for space reasons.

The above query $Q$ entails a {\em relevant rewriting}, according to the 
next definition.
\vspace{-2mm}
\begin{defn}[Relevant Rewriting of a Query]
Given a query $Q$ relevant to a mapping $\M_{ij} = S_i \rightarrow S_j$, 
its translation $Q^t$ is a relevant rewriting of $Q$ against $S_j$. 
We say that $\M_{ij}$ rewrites $Q$ into $Q^t$, denoted by $Q \stackrel{\M_{ij}}{\rightarrow} Q^t$.
\end{defn} 

Based on the above definition, we can now define a \emph{rewriting sequence}, as follows.

\vspace{-2mm}
\begin{defn}[Rewriting Sequence]
For a query $Q$, if $Q_0 \stackrel{\M_{01}}{\rightarrow} {Q_1}^t \stackrel{\M_{12}}{\rightarrow} {Q_2}^t 
\cdots \stackrel{\M_{(n-1)n}}{\rightarrow} {Q_n}^t$, we say that $Q$ 
rewrites into ${Q_n}^t$. The mappings $\M_{01}, \cdots, \M_{(n-1)n}$
are called the rewriting sequence.
\end{defn} 

\begin{figure}[t!]
\begin{center}
\includegraphics[width=6cm]{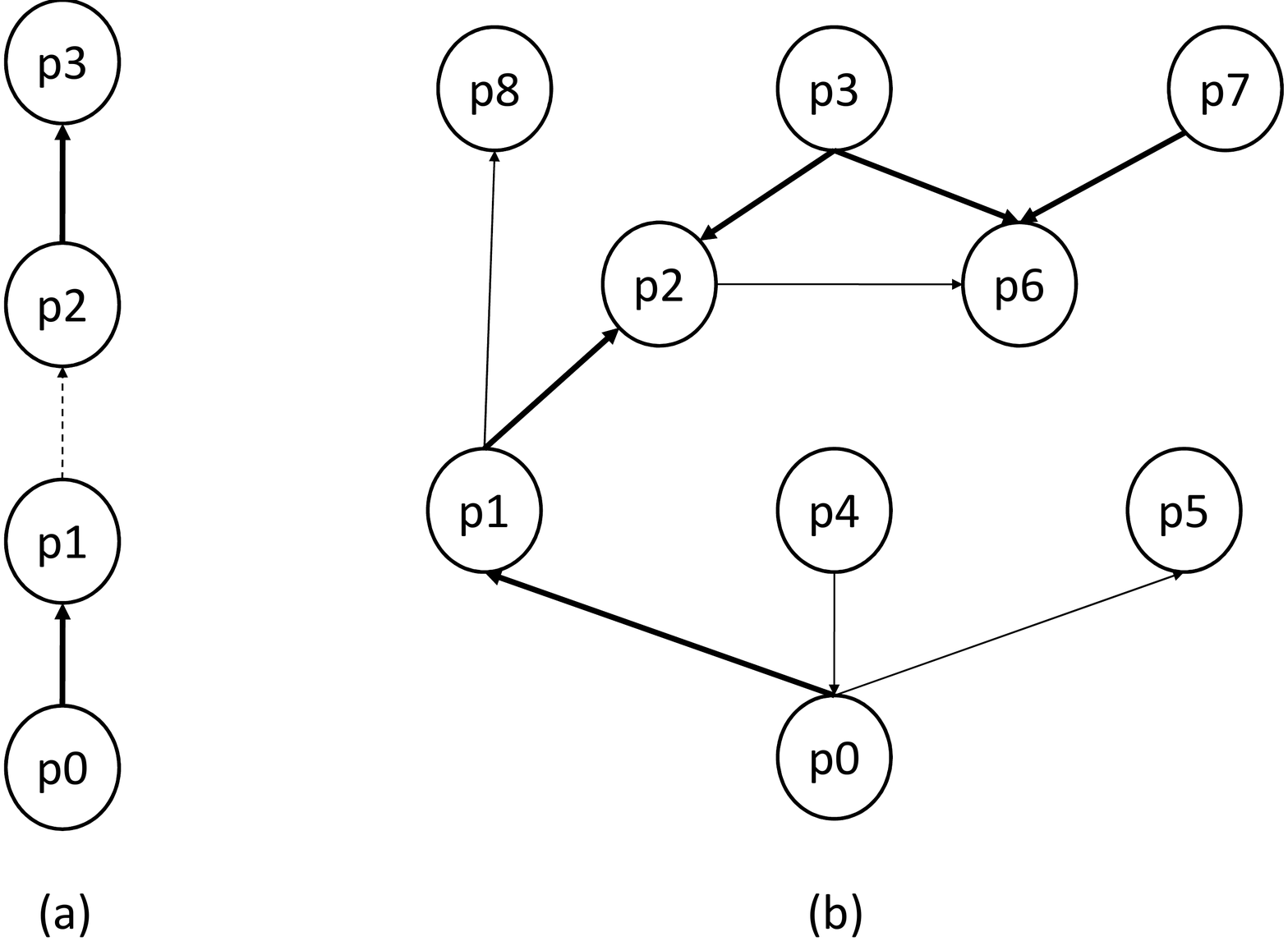}
\caption{(a) Useless rewriting sequence, (b) alternative rewriting sequences and relevant mappings (in bold).}
\label{fig:mapping-path}
\end{center}
\end{figure}

An example of \emph{rewriting sequence} starting from the peer $p_0$ to the peer $p_7$ is highlighted in bold in Figure~\ref{fig:mapping-path} (b). 

According to our problem definition, we need to find all the possible rewriting sequences of 
a given input query $Q_0$ on the initiating peer $p_0$. 
However, a rewriting sequence might not always exist between 
$p_0$ and an arbitrary peer $p_n$, since there might be an intermediate mapping that does not entail a relevant rewriting of the query. 
We denote such mapping as a \emph{useless mapping}
and the entire sequence a \emph{useless rewriting sequence}. An example of such a sequence is 
depicted in Figure~\ref{fig:mapping-path} (a), from $p_0$ to $p_3$, where the mapping from peer $p_1$ to $p_2$ is not relevant.   
Avoiding useless sequences is quite straightforward because they can be detected by adopting a local metric to assess whether the target of the current peer is able to handle the query, before actually shipping the query itself to 
that target. 
Such evaluation can be done by using the mapping rules themselves, as they are locally stored on the current peer and can be easily inquired to that purpose. 

Another issue that often occurs is that of alternative rewriting sequences, as depicted in Figure~\ref{fig:mapping-path} (b). Indeed, the current peer may have multiple alternative paths to rewrite a given query, and may have to choose the most appropriate one. E.g. in Figure~\ref{fig:mapping-path} (b), $p_0$ could choose among three possible alternatives $p_1$, $p_4$ and $p_5$. Exhaustively pursuing all possible rewritings is obviously not efficient, due to the great number of destination peers and rewriting sequences. Moreover, only fews rewritings along the sequences may happen to be the most relevant ones, which is always preferable to pursue. 
To this purpose, we need a \emph{relevance score} for each possible rewriting sequence (described next) in order to be able to rank the possible rewriting alternatives.
Consequently, it becomes feasible to rewrite the queries along the most relevant paths (e.g. represented by the bold arrows in Figure~\ref{fig:mapping-path} (b)). 

\noindent {\em Remark.} We observe that one could apply Definition 4 in a straightfoward manner to address the previous problems.
However, a relevance score solely based on a local metric would not be sufficient as it would only check one mapping at a time. Conversely, one needs to check an entire rewriting sequence among the possible alternatives. Thus, a global metric is needed to assess the relevance of queries with respect to the mappings in a rewriting sequence.






\subsection{Relevance metric}

In this section, we present our novel relevance metric to quantify the 
degree of relevance of mappings in the network and the data structures 
that allow computing it.

\subsubsection{AF-IMF metric}
Our metric which we call AF-IMF, i.e. \emph{atom frequency, inverse mapping frequence}, is an adaptation of the classical information retrieval metric TF-IDF to schema mapping. 
Variations of the TF-IDF weighting scheme are often used by search engines as a central tool for scoring and ranking a document's relevance given a user query.
Similarly, AF-IMF is a statistical measure to evaluate how important a query atom is to a mapping in the entire collection. The importance increases proportionally to the number of times an atom appears in the mapping but is offset by the frequency of the atom in the collection. 

In the following, we first define the AF-IMF for an individual mapping rule, then we extend it to entire mappings.

We introduce the \emph{atom count} in the given mapping rule $m$, as the number of times a given query atom $a_q$ fully appears in $m$ by using constant and variable unifications. 
This count is usually normalized by the number of occurrences of all atoms in $m$.
We assume that each atom can only appear once in a mapping rule,
 thus implying that the atom frequency can be approximated to $1$. 

\[
\begin{array}{l}

    \mathrm{AF_{i,j}} = \frac{n_{i,j}}{\sum_k n_{k,j}} \simeq \frac{1}{k}\\

\end{array}
\]

where $n_{i,j}$ is the number of occurrences of the considered atom $a_i$ in $m_j$, and the denominator is the sum of number of occurrences of all $k$ atoms occurring in the body (head, resp.) of $m_j$, where $a_i$ respectively appears.
Note that having two separate AF on the body and head according to where the atom $a_i$ appears in the mapping rule $m_j$ 
is crucial to characterize the forward from backward relevance, respectively. 

The inverse mapping frequency is a measure of the general importance of the atom, obtained by dividing the total number of mapping rules by the number of mapping rules containing the atom in the body (head, resp.), and then taking the logarithm of that quotient.

\[
\begin{array}{l}

    \mathrm{IMF_{i}} = \log \frac{| M |}{|\{m_j: a_{i} \in m_j\}|}

\end{array}
\]

with $| M | = | \cup_{i=1 \dots n}{\mu_{st}} |$ being the total number of mapping rules in the network, which amounts to the union (without duplicates) of all the source-to-target tgds; and $|\{m_j : a_{i} \in m_j\}|$ being the number of mapping rules where the atom $a_{i}$ appears (that is $n_{i,j} \neq 0$) in the body (head, resp.). If the atom is not in the network, this will lead to a division-by-zero, thus it is common to use $1$ + $|\{m_j : a_{i} \in m_j\}|$ instead.

Notice that the computation of AF depends on both the current query atom $a_i$ and the current mapping rule $m_j$. Differently, the IMF computation does not depend on the current mapping rule $m_j$ but only on the current query atom $a_i$.

Then,
\[
\begin{array}{l}

    \mathrm{(AF\mbox{-}IMF)_{i,j}} = \mathrm{AF_{i,j}} \times \mathrm{IMF_{i}} \simeq \frac{1}{k} \times \mathrm{IMF_{i}}

\end{array}
\]

The above formula implies that the mapping rules with less atoms are preferred with 
respect to those with more atoms. Therefore, a high weight in AF-IMF is reached by 
mapping rules with low total number of atoms, and low frequency in the 
global collection of mapping rules. 

What has been already observed above on forward from backward relevance implies that
a different value of the AF-IMF is computed for atoms appearing in the 
body (head, resp.) of the mapping rules in a similar fashion. 

A further step would lead to extend the above metric for the query atoms $a_q$ altogether so that it is possible to assign a comprehensive value of relevance the entire query $Q$ with respect to the mapping rule $m_j$ (as opposed to the previous case, when only an individual query atom $a_i$ was considered). Such step implies a simple measure (e.g. the {\em sum}) to put together the AF-IMF scores separately obtained by the query atoms $a_q$ of $Q$.

After applying the composition of the above scores, we obtain the overall score for the mapping rule $m_j$, as in the following:

\[
\begin{array}{l}
    \mathrm{(AF\mbox{-}IMF)_{j}} = \sum_i \mathrm{(AF\mbox{-}IMF)_{i,j}}
\end{array}
\]

After defining the notion of AF-IMF for an individual mapping rule $m$, we now extend the definition to the entire mapping $\scenario{M}$. We recall that the final goal of our metric is to assign a relevance value to those mappings that the current peer is about to evaluate in order to realize the query translation of $Q$.

Being a mapping scenario $\scenario{M} = (\schema{S}, \schema{T}, \mu_{st})$ defined by means of a set of $K$ ($K > 0$) mapping rules in $\mu_{st}$, we compute the overall AF-IMF score for $\scenario{M}$ as the sum of the AF-IMF scores obtained by each mapping rule $m \in \mu_{st}$ (according to the forward or backward definition of relevance). 

In other words, if the relevance is backward the query $Q$ matches the head side of the mapping rule $m_j$ (see Definition 3), the AF-IMF computation is done 
as shown below:

\[
\begin{array}{l}

    \mathrm{(AF\mbox{-}IMF)_{\scenario{M},head} = \sum\limits_{j=1}^{K_h} (AF\mbox{-}IMF)_{j}} \\

\end{array}
\]

where $K_h \subseteq K$ is the number of rules $m_j \in \mu_{st}$, such that 
$Q$ matches their head side. 

Instead, if the relevance is forward the query $Q$ matches the body side of the mapping rule $m_j$ (see Definition 2), the AF-IMF computation is done as shown below:
\[
\begin{array}{l}

    \mathrm{(AF\mbox{-}IMF)_{\scenario{M},body} = \sum\limits_{j=1}^{K_b} (AF\mbox{-}IMF)_{j}} \\

\end{array}
\]

where $K_b \subseteq K$ is the number of rules $m_j \in \mu_{st}$, such that
$Q$ matches their body side. 

The overall relevance of the query $Q$ with respect to the entire mapping $\scenario{M}$ is the maximum value between the two formulas above:
\[
\begin{array}{l}

    \mathrm{(AF\mbox{-}IMF)_{\scenario{M}} = max((AF\mbox{-}IMF)_{\scenario{M},head},(AF\mbox{-}IMF)_{\scenario{M},body})} \\

\end{array}
\]

In such a way, given a query $Q$ as input, the AF-IMF metric assigns a score of relevance to each inward and outward mapping of the peer, to let it choose the most relevant paths for query translation, i.e. the ones with the highest scores.

\begin{example}
Consider Figure~\ref{fig:schemaMappingExample2} that is a slightly different version of Figure~\ref{fig:schemaMappingExample}.
A set of correspondences $v_1, v_2, v_3, v_4$ and $v_5$ connects elements in the two schemas. 

\begin{figure}[t!]
\begin{center}
\includegraphics[width=8cm]{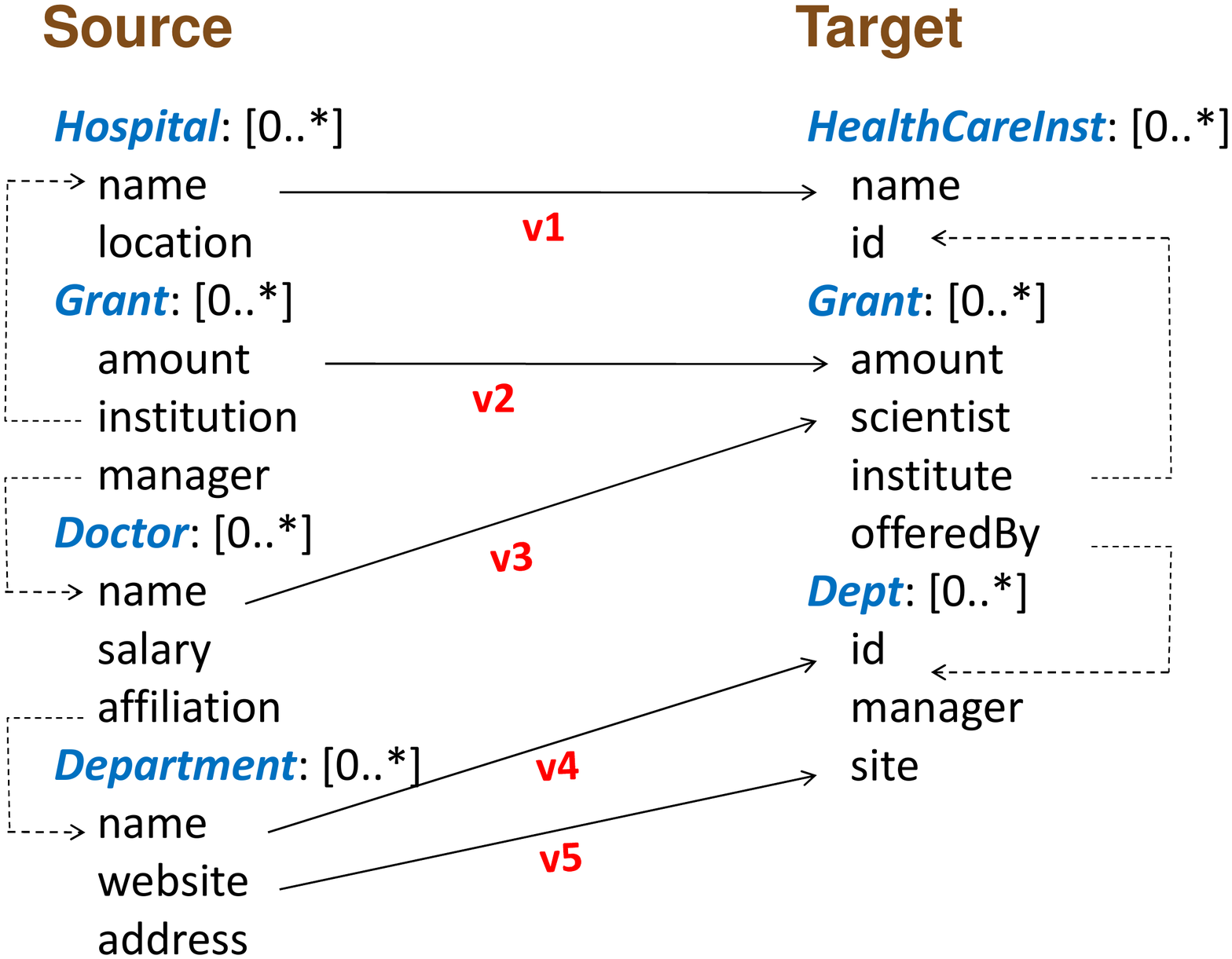}
\caption{A Schema Mapping Example}
\label{fig:schemaMappingExample2}
\end{center}
\end{figure}

\noindent Assume that the set of corresponding s-t tgds is the one reported below:

\removesmallspace
\[
\begin{array}{l}
\textsc{Source-to-Target Tgds} \\
m_1.\tld \forall n, l\!: \emph{Hospital}(n, l) \rightarrow \exists I\!\!: \emph{HealthCareInstitution}(n, I) \\
m_2.\tld \forall n, s, d, dw, da, a, pi, l\!: \emph{Doctor}(n, s, d) \wedge \emph{Grant}(a, pi, n)\\
\quad \quad   \wedge \emph{Department}(d, dw, da) \wedge \emph{Hospital}(pi, l) \\
\quad \quad   \rightarrow \exists I,g\!\!: \emph{HealthCareInstitution}(pi, I) \wedge \emph{Grant}(a, n, I, d) \\ 
\quad \quad   \wedge \emph{Dept}(d, g, dw)\\
m_3.\tld \forall n, w, a\!: \emph{Department}(n, w, a) \rightarrow \exists g\!\!: \emph{Dept}(n, g, w) \\
\end{array}
\removesmallspace
\]

\noindent The mapping $\M$ among source $S$ and target $T$ includes all the above three mapping rules.

Now, let us imagine that the source peer $S$ is connected to other target peers ($T_1$, $T_2$ and $T_3$) all having, for simplicity, an identical target schema $T$ with sets of different mappings. Such mappings ($\M_1$, $\M_2$ and $\M_3$) are simply variants of $\M$, i.e. mappings derived from $\M$ by including a different subset of the mapping rules of $\mu_{st}$, as specified in the following:
\begin{itemize}
    \item $\M_1$ : $\mu_{st} = \{m_1, m_2\}$
    \item $\M_2$ : $\mu_{st} = \{m_1, m_3\}$
    \item $\M_3$ : $\mu_{st} = \{m_2, m_3\}$
\end{itemize}

If we compute the AF-IMF scores for all the mappings above, i.e. $\M$, $\M_1$, $\M_2$ and $\M_3$, it is easy to check that $\M$ will always get the highest score, since it is the most complete mapping. Therefore the peer T, that is connected to S through $\M$, represents the most relevant peer to follow in the query reformulation process.
\end{example}

Nevertheless, a further complication arises since IMF cannot be exactly computed as the size of the entire collection of
mapping rules at a given time is not known, due to the fact that the network is dynamically 
changing.  

To address this problem, each peer is equipped with a 
set of semantic data structures, that summarizes the local and external 
mappings of a peer (see next Section for details).
Thus, by exploiting such data structures, we can compute an approximation of IMF
for the distributed case, as discussed in Section~\ref{sect:how-to}.

\subsubsection{Semantic Data Structures}
\label{sec:datastruct}
In this section, we first introduce the local semantic data structures stored on  
each peer. Then, in Section~\ref{sect:how-to}, we present 
how they can be exploited to approximate the IMF values.

\begin{figure}[t!]
\vspace{-2mm}
\begin{center}
\includegraphics[angle=-90,width=9cm]{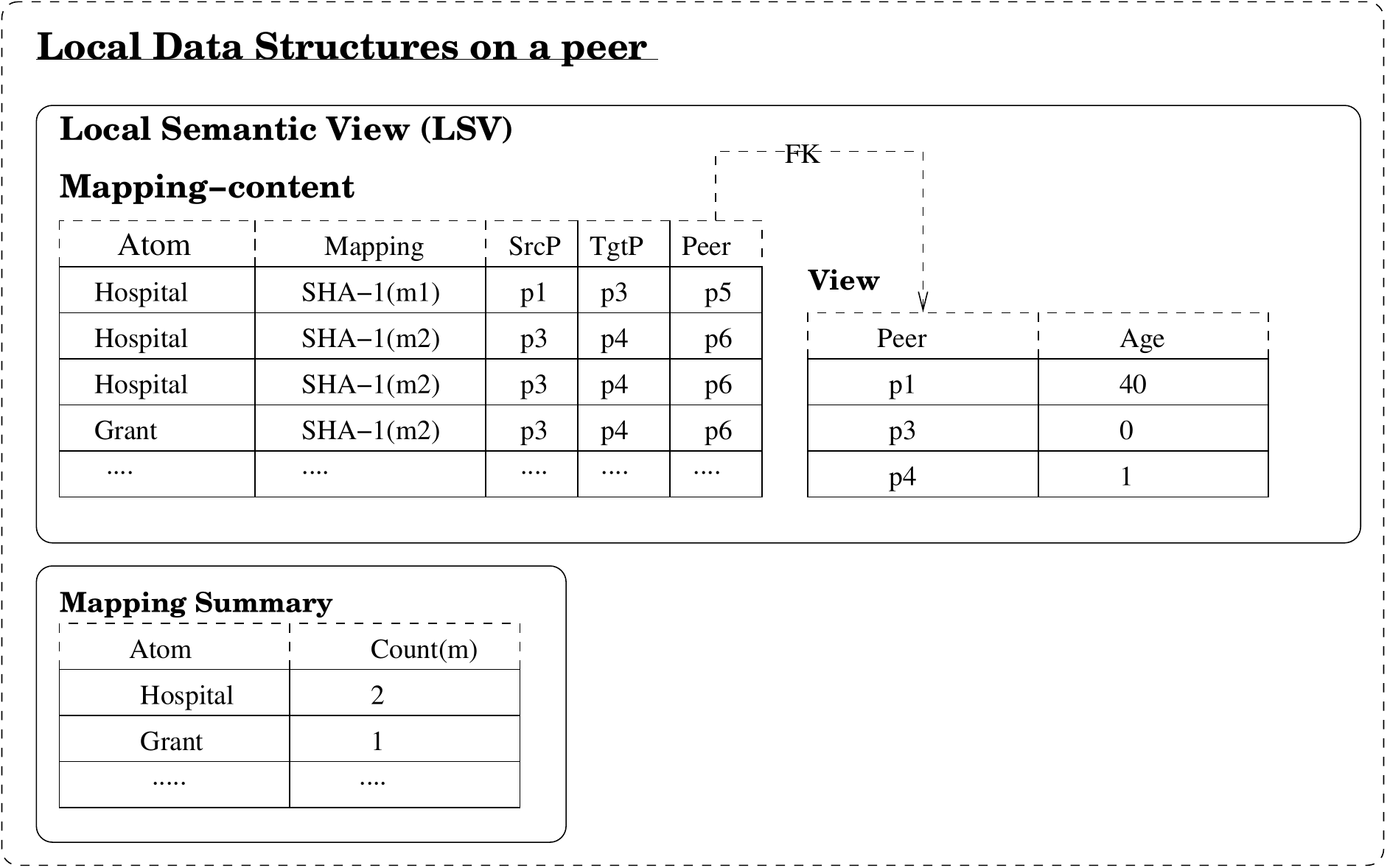}
\vspace{-2mm}
\caption{Local data structures on a peer.}
\label{fig:viewExample}
\end{center}
\vspace{-5mm}
\end{figure}

Figure~\ref{fig:viewExample} represents the local data structures
on each peer.
Each peer maintains a set of local or internal mapping rules\footnote{The mapping statements have been omitted from Figure~\ref{fig:viewExample} to avoid clutter.}, i.e. mapping rules  
from its local schema to the schema of each of its acquaintances, the latter being a selected subset of the peer's neighbors~\cite{miller03,Halevy05}. 
Moreover, it also stores a \emph{Local Semantic View} (LSV in short), 
that encloses information about external mapping rules (distinct from the local ones), selected uniformly at random from the network. 
This view is used to compute the relevance values. 
Precisely, an LSV for each peer consists of: a five-column table \emph{Mapping-content} $(Atom,\ Mapping,\ SrcPeer,\ TgtPeer,\ Peer)$, and of a two-column table \emph{View} $(Peer,$ $Age)$, with a foreign key 
constraint between \emph{View.Peer} and \emph{Mapping-content.Peer}.  
The \emph{Map\-ping-content} relation has a column \emph{Atom} containing the atom of a mapping rule 
in the network; a column \emph{Mapping} containing the ID of the mapping rule in which  \emph{Atom} appears; a column \emph{SrcPeer} containing the ID of the external source peer to which \emph{Mapping} is an outward mapping; a column \emph{TgtPeer} containing the ID of the external target peer to which \emph{Mapping} is an inward mapping; a column \emph{Peer} containing the ID of the peer in the network that has provided the current tuple in a gossip cycle.
The \emph{View} relation has a column \emph{Peer} containing the ID of a peer in the network; a column \emph{Age} containing a numeric field that denotes the age of the mapping rules since the time in which they have been included within the \emph{View}. Figure~\ref{fig:viewExample} shows an example of a LSV on a peer. 

To uniquely identify each mapping rule in the PDMS, we assign an ID 
to each mapping, using cryptographic hash functions (e.g. SHA-1) to reduce the probability of 
collision~\footnote{In DHTs or structured P2P networks, on which PDMS are based, a unique key identifier is assigned to each peer and object.
IDs associated with objects are mapped through the DHT protocol to 
the peer responsible for that object. In our setting, each object is a mapping.}.

As the size of the LSV is limited, it implies that the view entries need to 
be replaced, based on their age information. 
In order to maintain each LSV on the peers, we adopt classical thread-based gossiping mechanisms, 
aiming at updating the LSV with newly incoming tuples from the outside. 
In Section 4.4 we provide the details of such maintenance. 





Besides local mappings, each peer also maintains
 an additional descriptive data structure of such mappings, called \emph{Mapping Summary}, which is implemented as a local 
Bloom filter~\cite{Bloom70space}.
A Bloom filter is a method for representing a set $A = \{a_1, a_2, \cdots, a_n\}$
of $n$ elements (also called keys) to check the membership of any element in $A$.  
In the \emph{Mapping Summary}, a bit vector $v$ of $m$ bits, initially set to $0$, 
represents the positions of $k$ independent hash functions, $h_1, h_2, \cdots, h_k$,
each with range $\{1, \cdots, m\}$. In the \emph{Mapping Summary}, a key is built as follows: 
for each mapping rule, the conjunction of all atoms in the 
body $\phi$ (in the head $\psi$, resp.) is a key; 
each individual atom $a_i$ in the 
body $\phi$ (in the head $\psi$, resp.) is a key; 
each subset $a_1, \dots, a_n$ of the atoms in the 
body $\phi$ (in the head $\psi$, resp.), such that it exists at least one joined variable in each atom $a_i$
and $a_{i+1}$, is a key. 
By enumerating the above keys, each body (head, resp.) of the mapping rule has a total of 
$\frac{n(n+1)}{2}$
entries in a \emph{Mapping Summary}.
Such a combination of atoms and/or individual atoms may 
appear in several distinct local mapping rules on that peer. For each atom (or combination of atoms thereof) $a \in A$, the bits at positions $h_1(a), h_2(a), \cdots, h_k(a)$ in $v$ are set to $1$.     
A membership query checks the bits at the positions $h_1(a), h_2(a), \cdots, h_k(a)$. 
If any of them is $0$, the atom $a$ is not in the set. Otherwise, we conjecture that 
$a$ is in the set, although this may lead to a false positive. The aim is 
to tune $k$ and $m$ so as to have an acceptable probability of false positives.  
The advantage of using Bloom filters resides in the fact that they require 
very little storage, at the slight risk of false positives. Such 
probability is quite small already for a total of $4$ different hash functions~\cite{fan98}. Figure~\ref{fig:viewExample} shows an example of a \emph{Mapping Summary} on a peer. 

Similarly to the LSV, the \emph{Mapping Summary} needs to be maintained in the presence of changes of the atoms within the mapping rules, and/or additions and deletions
of the mapping rules themselves.
This is done by maintaining in each location $l$
in the bit vector $v$, a count $c(l)$ of the number of times that the bit is set to 
$1$. The counts are initially all set to $0$. When insertions or deletions take place, 
the counts are incremented or decremented accordingly. 




Finally, to allow friendship linking among peers, a peer mantains a third structure, that is basically a local \emph{FOAF file} containing the URIs of its friends FOAF files. 
Whenever a user (or a peer) generates its FOAF file, it can obtain an identity for that file in the form of a URI. This URI could point to a reference in a friend's FOAF file.
URIs correspond to unique peer and object identifiers in a PDMS.  
In particular, a peer $p_1$ may need to store into its FOAF file:
(1) the list of other peers he knows and he is friend with, as a link to its
friend's FOAF file (e.g. P3.rdf in the example); (2) possibly, the link to its friend's \emph{Mapping
Summary (e.g. $P3\_MapSum$ in the example below)}. 

\begin{scriptsize}
{\bf
\begin{verbatim}

<foaf:knows>
  <foaf:Peer>
    <foaf:peerID> P3</foaf:peerID>
         <rdfs: seeAlso rdf: resource = 
                `http://www.mirospthree.com/P3.rdf'/>
         <rdfs: seeAlso rdf: resource = 
                `http://www.mirospthree.com/P3_MapSum'/>
  </foaf:Peer>
</foaf:knows>

\end{verbatim}
}
\end{scriptsize}

The main goal of FOAF files is to maintain the current friendship links 
of a given peer. During query translation, 
the FOAF file is expanded by adding new friends, by invoking the Algorithm FindDirectFOAFFriends, 
described in Section~\ref{sec:algos}. 
Notice that adopting and exploiting the friendship links of a given peer during the query translation process is complementary to exploiting the semantic mappings towards the peer's acquaintances. In fact, the friendship links are especially useful in the presence of network churn, as they act as a background network regardless of the peer's acquaintances and its direct inward/outward mappings. A more detailed experiment about network churn, scalability and the usefulness of FOAF links 
is provided in Section~\ref{sec:expe}. 

In our model, no peer can access the other peer's mapping summary until an 
explicit friendship link has been established between such peers, 
thus leading to modify their respective FOAF files accordingly.
This mechanism gracefully replaces an explicit negotiation and coordination among peers for accessing 
their respective data structures. An additional access control mechanism, e.g.~\cite{ang10},
can be adopted on top of FOAF files to further strenghten the security of the 
network. 

In the remainder of this discussion and in Section~\ref{sec:algos}, 
we denote the peers indexed in a FOAF file as `friends'.
These represent the peers whose mapping summary can be accessed, in order to  
widen the scope of the queries. In particular, in Section~\ref{sec:algos},
we will discuss how to enlarge the set of simple friends of a peer by 
exploiting friendship links in its FOAF file.

\subsubsection{Distributed computation of AF-IMF}~\label{sect:how-to}
Using the local semantic view and the local mapping rules, we can compute $IMF_i$ distributively, as follows. 
Let $k$ be the number of distinct local mapping rules entries and let $t$ the
number of distinct mapping rules in the LSV. We know by definition that the 
$k$ entries and $t$ entries are not overlapping, thus we may say that locally we 
have $k + t$ mapping rules. 
Then, we have to determine what is the approximation of $|M|$, the total number of 
mapping rules in the collection, possibly without duplicates. We may think of computing $N$, the total number of peers in the network and multiplying it by $k + t$, thus obtaining $|M| = 
(k + t) \times N$. Moreover, we observe that $N$ can be easily computed if we know the network topology. 
For instance, for DHTs it suffices to record the size of the routing table, which is 
$r = log(N)$, and by taking the inverse as $2^r = N$. For super-peer networks, we may 
have an entry point that registers the total number of peers $N$. For 
unstructured P2P networks, we may rely on flooding to count the total number of 
peers in the network.
In a similar way, the $|\{m_{ij} : a_{i} \in m_{ij}\}|$ can be computed by 
selecting among the $k$ and $t$ local mappings, those that contain the
atom $a_i$, thus obtaining $|\{m_{ij} : a_{i} \in m_{ij}\}| = 
(k_i + t_i) \times N$.  

\emph{However, we need to avoid duplicate mapping rules in the previous computation.}
In order to do this, we need to uniquely identify a mapping in the entire network.
A simple and effective way to do this is to couple each mapping with its  
signature, using a cryptographic hash function (e.g. SHA-1). 
We present in Section~\ref{sec:algos} an algorithm to compute 
AF-IMF distributively, that avoids duplicate mappings by using signatures.   

\noindent {\em Remark.} 
As a final observation, we underline that the problems illustrated in Figure~\ref{fig:mapping-path}
are both overcome, since the useless sequences do not affect the AF-IMF metric. Moreover, AF-IMF enables the search of the most relevant rewriting sequences in a global fashion, as expected by our previous reasoning. 
In the experimental analysis (Section~\ref{sec:expe}), we show the effectiveness of this metric, also
when compared to a local metric (e.g. by adopting the sole AF as a local metric).

 



\section{Reformulation Algorithms}~\label{sec:algos}
In this section, we illustrate our query reformulation algorithm: the core algorithm that 
translates a query based on relevance; an algorithm for seeking new friends that contain relevant mappings for the query; a distributed algorithm to compute the relevance of mappings, that is used by the two former algorithms. Finally, we briefly discuss the gossiping algorithm for updating semantic views.



\subsection{Distributed computation of the relevance} 

Algorithm~\ref{algo:compute-relevance} computes a measure of the relevance of a set of mapping
rules on a given peer
with respect to an input query, with the aim of getting an ordered top-k list of mappings
to be exploited (by Algorithm~\ref{algo:query-map}) and the aim of finding new friends by (Algorithm~\ref{algo:find-peer}).

The algorithm has two main parts. Lines 1-14 aim at computing the 
IMF values for each query atom, and this entails a separate computation, depending on which 
side of the mapping rule the query atom belongs.   
Therefore, two vectors $BodyIMF$ and $HeadIMF$ are built to store the IMF values of each atom in the query $Q$.

Then, the second part of the Algorithm (lines 15-32) 
computes the AF values by counting the number of times that a 
query atom occurs in the matched side of the mapping rule, and the complete value of 
AF-IMF is then returned. 
The final relevance value (line 32) for the whole mapping rule is taken by applying a suitable ranking function to the values in the 
above vectors (e.g. {\emph sum}). 

Let us observe that the computation of the IMF 
only depends on the atom $a_i$ in the query $Q$, and not on the current mapping rule. 
For this reason, we also make sure that the computation at lines 1-14 is done only once for the same query, 
by saving intermediate results.  

Indeed, the computation is done by asking each known peer (both destination peers through mappings and 
new discovered peer friends in the FOAF file $f$).
The maximum number of inquiries is given by the $REQS$ threshold. Observe that if $REQS = 0$ no external inquiries have been done, and only the entries of the current peer's LSV have been inspected, whereas 
a value of $REQS$ greater than $0$ leads to also inspect the LSV of external peers. 
Also note that such inquiries are done by discarding duplicates through the asynchronous method GetDistinctMappingRules, 
that checks the signatures of the mapping rules. We omit the pseudo-code of this method for space reasons. 

\begin{figure}[t!]
\begin{center}
\includegraphics[width=7cm]{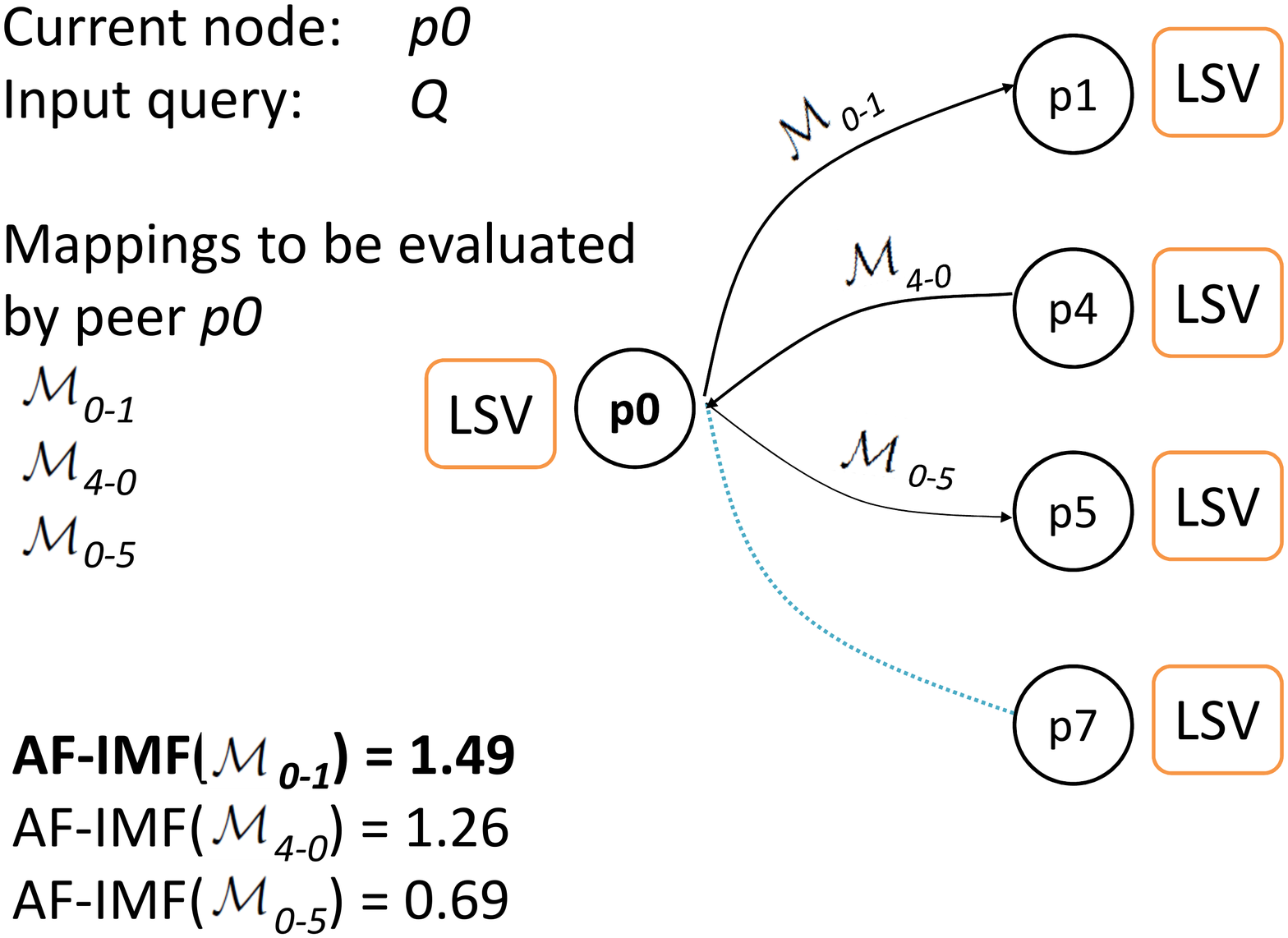}
\caption{An example of ComputeRelevance (Algoritm 1).}
\label{fig:Relevance}
\end{center}
\end{figure}

Figure~\ref{fig:Relevance} shows an example of how Algorithm~\ref{algo:compute-relevance} computes the relevance. A query $Q$ is initially posed against the peer $p_0$, which in turn chooses among three alternative target peers (also called acquaintances). Also, note that from $p_0$ toward $p_7$ there is no direct mapping, but rather a FOAF link depicted by a dotted blue arrow. 
Thus, mappings $\M_{01}$ (from $p_0$ to $p_1$), $\M_{40}$ (from $p_4$ to $p_0$) and $\M_{05}$ (from $p_0$ to $p_5$) must be evaluated aiming at finding the \emph{top-k} relevant ones for the input query (in this example, we assume for simplicity that \emph{k = 1}). By inspecting $p_0$'s LSV, Algorithm~\ref{algo:compute-relevance} performs the computation of the relevance metric for each mapping rule $m$ of each mapping involved ($\M_{01}, \M_{40}, \M_{05}$), by assigning an AF-IMF value to each involved atom, as previously discussed.
At the end, the mapping $\M_{01}$ (from $p_0$ to $p_1$) gets the highest relevance score amongst all the other mappings, thus becoming the \emph{top-1} step in the rewriting sequence of query $Q$.  

\begin{algorithm}[t!]
\SetAlFnt{\scriptsize\sf} 
\AlFnt{\scriptsize\sf} 
\linesnumbered
\SetKwInOut{Input}{Input}
\SetKwInOut{Output}{Output}
\SetKwComment{Comment}{//}{}
\caption{{\bf ComputeRelevance} - computes the relevance according to the gossiped information in the local semantic view\label{algo:compute-relevance}}
\Input{A query $Q$ as set of atoms ${\cal A}_{Q}$, a list of k mapping rules $m_k$, 
a peer $p$ with its LSV and FOAF file $f$}
\Output{The vector of relevance values $RV$ for the input list of k mapping rules}
\ForEach{atom $a_i$ in ${\cal A}_{Q}$}{
		\Comment{Compute the IMF value according to the $matchedSide$}
		$n$ = total nr. of mapping rules in the LSV\;
		$nB_i$ = total nr. of mapping rules in the LSV containing $a_i$ in the body\;
		$nH_i$ = total nr. of mapping rules in the LSV containing $a_i$ in the head\;
		\textbf{Let} $count_{reqs} = 0$\; 
		\ForEach{$p'$ in the View of LSV and in the FOAF file $f$}{
			\If{$count_{reqs} >= REQS$}{
				break\;
			}
			$n$ += \textbf{GetDistinctMappingRules}($p'$)\;
			$nB_i$ += \textbf{GetDistinctMappingRules}($p'$, "Body", $a_i$)\;
			$nH_i$ += \textbf{GetDistinctMappingRules}($p'$, "Head", $a_i$)\;
			$count_{reqs}$++\;
		}
		$BodyIMF[i]$ = log($n$ / (1 + $nB_i$))\;
		$HeadIMF[i]$ = log($n$ / (1 + $nH_i$))\;
}
\ForEach{mapping rule $m_k$ in the list of input mapping rules}{
	\eIf{all atoms $a_i$ in ${\cal A}_{Q}$ are in the body of $m_k$}{
		$matchedSide$ = "Body"\;
	}{
		\eIf{all atoms $a_i$ in ${\cal A}_{Q}$ are in the head of $\mu_k$}{
			$matchedSide$ = "Head"\;
		}{
            \Comment{No relevance}
			$RV[k]$ = 0\;	
			continue\;
		}
	}
	\ForEach{atom $a_i$ in ${\cal A}_{Q}$}{
		$AF-IMF[i]$ = $0$\;
		\Comment{Compute the AF-IMF value for $a_i$ according to the $matchedSide$}
		\eIf{$matchedSide$ == "Body"}{
            $BodyAF_i$ = count of the nr. of $a_i$ in the body of $m_k$\\
			$AF-IMF[i]$ = $BodyAF_i$ * $BodyIMF[i]$\;
		}{
            $HeadAF_i$ = count of the nr. of $a_i$ in the head of $m_k$\\
			$AF-IMF[i]$ = $HeadAF_i$ * $HeadIMF[i]$\;
		}
	}
	\Comment{Compute the final relevance value $RV$ for the whole mapping rule $m_k$}
	$RV[k]$ = RankFn($AF-IMF[i]$)\\
}
\textbf{return} $RV$\;
\end{algorithm}
\vspace{-1mm}

\subsection{Translating queries based on relevance}

\begin{figure}[t!]
\begin{center}
\includegraphics[width=7cm]{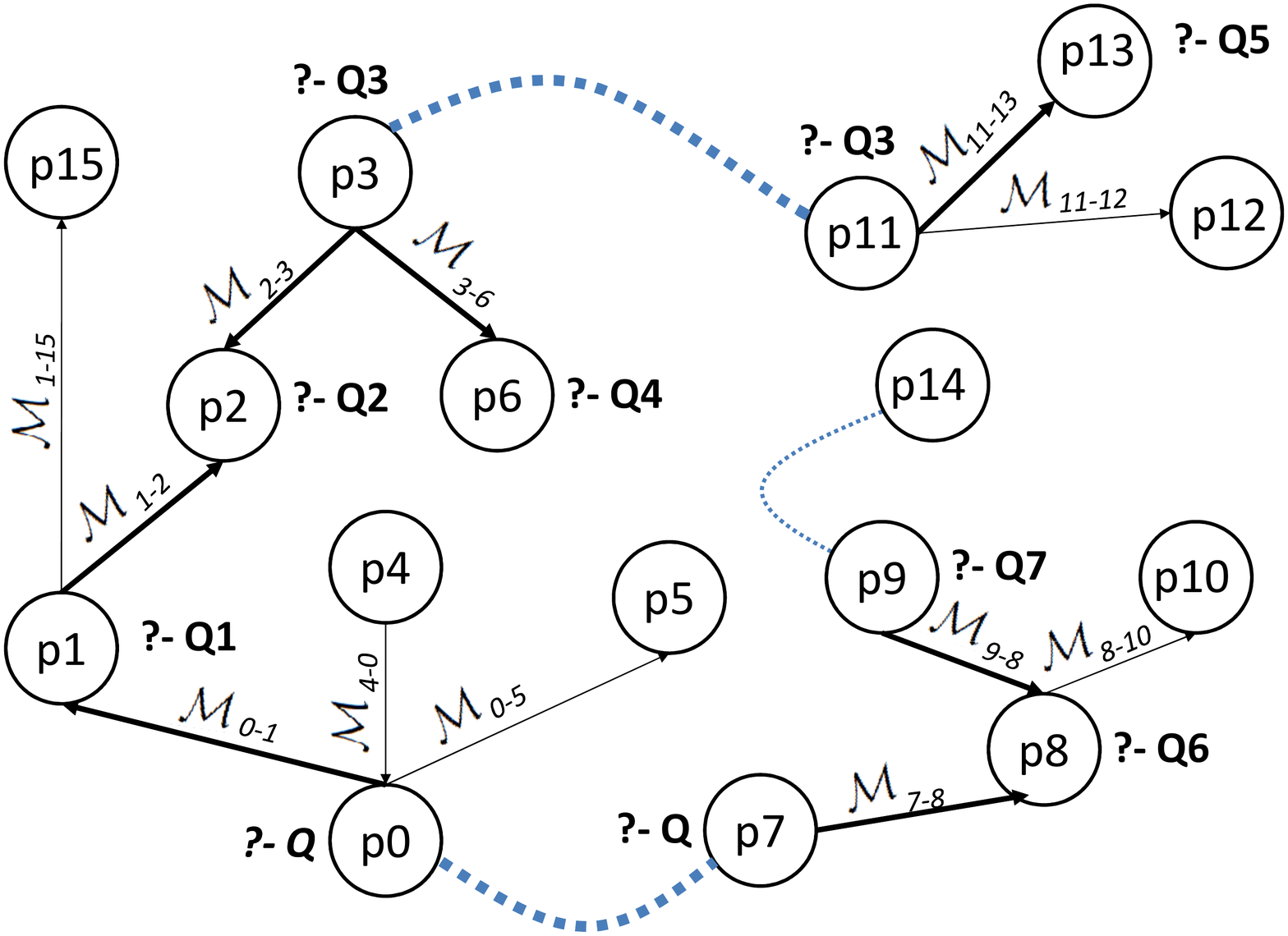}
\caption{An example of QueryTranslate (Algorithm 2).}
\label{fig:Translate}
\end{center}
\end{figure}

\begin{algorithm}[t!]
\SetAlFnt{\scriptsize\sf} 
\AlFnt{\scriptsize\sf} 
\linesnumbered
\SetKwInOut{Input}{Input}
\SetKwInOut{Output}{Output}
\SetKwComment{Comment}{//}{}
\caption{{\bf TranslateQuery} - Query translation based on relevance\label{algo:query-map}}
\Input{Query $Q$ as set of atoms ${\cal A}_{Q}$ and a peer $p$ with its list $MList$ of local mappings $\Sigma_{1\leq i\leq n}\M_{i}$, and its FOAF file $f$}
\Output{Query results $res$ of the query $Q$ against the peer $p$ exploiting both the set of local relevant mappings 
$\Sigma_{1\leq i\leq n}\mu_{i}$ and new relevant peer friends}

\If{$Q$.query-hops $\geq \alpha$}{
	return $res$\;
}
increase $Q$.query-hops by 1\;
Call \textbf{FindDirectFOAFFriends($Q$, $p$)}\;
\textbf{Let} $L$ be a list of mappings ordered by relevance\;
\ForEach{local mapping $\M_{i}$ in $MList$} {
    $RV$ = Call \textbf{ComputeRelevance}($Q$, $\M_{i}$.MappingRules())\;
    $mapscore[i]$ = SumValuesFromVector($RV$)\;
}    
$L$ = \textbf{Order} $MList$ according to mapping relevance values in $mapscore$\;
\ForEach{top-k ordered mapping $\M_{i}$ in $L$}{
	\If{$\M_{i}$ has been already processed}{
		continue\; \Comment{To avoid cycles}
	}
	\textbf{Let} $destPeer$ the destination peer through mapping $\M_{i}$\;
	\eIf{$Q$ is relevant to the body of $\M_{i}$}{
		Translate $Q$ along $\M_{i}$ obtaining $Q'$\\
		\eIf{$\M_{i}$ is outward}{
			$res$ = $res$ $\cup$ Eval($Q$)\;
			$res$ = $res$ $\cup$ TranslateQuery($Q'$, $destPeer$)\;
		}{
			$res$ = $res$ $\cup$ Eval($Q'$)\;
			$res$ = $res$ $\cup$ TranslateQuery($Q$, $destPeer$)\;
		}
	}{
	\If{$Q$ is relevant to the head of $\M_{i}$}{
		Translate $Q$ against $\M_{i}$ obtaining $Q'$\\
		\eIf{$\M_{i}$ is outward}{
			$res$ = $res$ $\cup$ Eval($Q'$)\;
			$res$ = $res$ $\cup$ TranslateQuery($Q$, $destPeer$)\;
		}{
			$res$ = $res$ $\cup$ Eval($Q$)\;
			$res$ = $res$ $\cup$ TranslateQuery($Q'$, $destPeer$)\;
		}
	}
	}
} 

\textbf{Let} $F$ be a list of friends\;
$F$ = Call \textbf{ComputeFriendsWithGreatestCount($Q$, $f$)}\;
\ForEach{top-k ordered friend $pFoaf$ in $F$}{
\Comment{To exploit new interesting peer friends}
		$res$ = $res$ $\cup$ TranslateQuery($Q$, $pFoaf$)\;

}
\textbf{return} $res$\;
\end{algorithm}
\vspace{-1mm}
Algorithm~\ref{algo:query-map} translates a query initiated at a peer, first against its set of local mappings and then by exploiting local friendship links at that peer. 
The algorithm is inherently recursive, and at each iteration increases the number of \emph{query 
hops}, until a given threshold $\alpha$ is reached. This avoids exploring the entire
network, by conveying the query toward a limited number of peers.
By exploiting the notion of relevance for the input query $Q$, new \emph{friends} are discovered and added to the
FOAF friend list.


By invoking the method FindDirectFOAFFriends (line 4), the current peer 
enlarges the list of its friends in its local FOAF file. 
Therefore, new relevant friends might be discovered, similarly to real-life friendship mechanisms, and to friend-bases 
game applications (e.g. Farmville) in modern social platforms (e.g. Facebook). 
Lines 6-8 invoke the method ComputeRelevance for each local mapping $\M_i$ of the peer, in order to get the relevance scores for such local mappings. Then, at line 9, the list of local mappings is ordered according to the the calculated relevance scores with respect to the input query $Q$.

Mapping identity is checked in lines 10-11, in order to avoid 
using the same mappings more than once in different iterations. 
The query rewriting proceeds by taking into account the direction of the mapping
(cfr. Definition 1) and then can take place {\em along} (line 12) or {\em against} 
(line 21) the mapping, thus obtaining the translated query $Q'$. 
Then, according to the type of the mapping considered - if inward or outward, the input 
query $Q$ and the translated one $Q'$ are executed against the current peer or instead used in the recursive call of the Algorithm.
Next, the query translation task is pushed towards the new interesting peer friends 
encoded in the FOAF file $f$ (lines 31-34). This search exploits 
the peer friends' \emph{Mapping Summary} 
to check whether there is a high number of mappings that contains 
atoms of the input query $Q$ (via the method ComputeFriendsWithGreatestCount).
Finally, all the query results $res$ are returned (line 34) as the union of all the results 
harvested throughout the recursive invocations of the algorithm.

Figure~\ref{fig:Translate} shows an example of execution of Algorithm~\ref{algo:query-map}. A query $Q$ is posed against the peer $p_0$. In trying to choose the most relevant rewriting sequence (lines 5-30), $p_0$ applies 
Algorithm~\ref{algo:compute-relevance} (for simplicity, we assume that \emph{top-k = 1}). This way, the query $Q$ is rewritten and traslated transitively until $p_6$ is reached. No translation is further possible, since 
$p_6$ is a terminal node. 
However, FOAF links found in line 4 of the Algorithm~\ref{algo:query-map} are also exploited in this example. Indeed, they allow to traverse disconnected subsets of the nodes in the graph of Figure~\ref{fig:Translate} (lines 31-34). If the friend reachable through the link is able to treat the query, the query can be further propagated to that friend and its subgraph. In the figure, one can see that $p_7$ and $p_{11}$ receive the queries $Q$ and $Q3$ respectively from $p_0$ and $p_3$. By contrary, $p_{14}$ is not able to treat the query that $p_9$ holds, thus such a query is not propagated further.
Obviously, each friend would further spread the query, thus increasing the total number of relevant rewritings.

The following proposition holds. 
\vspace{-1mm}
\begin{proposition}
If $|{\cal A}_{Q}|$ is the size (number of atoms) of an input query $Q$ and $|M_{r}|$ the number of the relevant mappings in the PDMS then  
the number of rewritings generated by $TranslateQuery$ is $O(|M_{r}|^{|{\cal A}_{Q}|})$.
\end{proposition}  
\vspace{-1mm}
\begin{algorithm}[t!]
\SetAlFnt{\scriptsize\sf} 
\AlFnt{\scriptsize\sf} 
\linesnumbered
\SetKwInOut{Input}{Input}
\SetKwInOut{Output}{Output}
\SetKwComment{Comment}{//}{}
\caption{{\bf FindDirectFOAFFriends} - Finds the top-k relevant "Simple Friends" and adds their entries in the FOAF file\label{algo:find-peer}}
\Input{A query $Q$ as set of atoms ${\cal A}_{Q}$ and a peer $p$ with its list $LSVList$ of mappings 
$\Sigma_{1\leq i\leq n}\M_{i}$ in the peer's local semantic view (LSV) and a FOAF file $f$}
\Output{The updated FOAF file $f$}
\textbf{Let} $L$ be a list of mappings ordered by relevance\;
\ForEach{mapping $\M_{i}$ in $LSVList$} {
    $RV$ = Call \textbf{ComputeRelevance}($Q$, $\M_{i}$.MappingRules())\;
    $mapscore[i]$ = SumValuesFromVector($RV$)\;
}    
$L$ = \textbf{Order} $LSVList$ according to mapping relevance values in $mapscore$\;
\ForEach{top-k mapping $\M_{i}$ in the ordered list $L$} {
	Let $p'$ the target peer through $\M_{i}$\;
	\If{$p'$ is not in the FOAF $f$}{
		Call \textbf{InvitePeer($p$, $p'$)}\;\Comment{Asynchronous method}
		\If{the previous invitation has been accepted}{
			Insert $p'$ in the FOAF file $f$\;
		}
	}
}
\textbf{return} the updated FOAF file $f$\;
\end{algorithm}
\vspace{-2mm}
\subsection{Seeking new friends} 
Algorithm~\ref{algo:find-peer} updates the FOAF file of a given peer, by adding new friends, discovered after 
an exhaustive inspection of the content of the local semantic view of a peer. 
Before adding a peer $p'$ to the FOAF file of the current peer, a formal invitation is sent and must be accepted. A simple extension of Algorithm~\ref{algo:find-peer} can be thought, in which an external peer, which is not 
friend of a friend, is added to the FOAF file. We omit its pseudocode for the sake of conciseness.

\subsection{Gossiping mapping entries}
To conclude this section, we discuss the gossip behavior of each peer. An active thread describes how a 
peer $p$ initiates a periodic gossip exchange, while the passive thread takes care of a 
gossip exchange initiated by some other content peer $p''$. 

The active behavior is triggered after each time interval $TGossip$.  After incrementing the 
age of its view entries by $1$, the peer $p$ selects from its view: (a) a peer $p'$, being the oldest 
contact via select\_oldest()  and (b) a viewSubset, being a random subset of Mapping-content within the local semantic 
view of $LGossip$ 
size (where $0 < LGossip  <= VGossip$). Then, peer $p$ send to $p'$ a gossip message, a message that contains the viewSubset. Recall that each peer keeps in its LSV a set of the mappings containing a 
specific atom (see Figure~\ref{fig:viewExample}). 

The peer $p$ receives in exchange $gossipMsg'$ containing similar information from $p'$, and
 creates a viewEntry related to $p'$, with age $0$. Next, peer $p$ discards 
 duplicates view entries through using Merge. This lets taking care of 
 the problem of redundant rewriting sequences.


The passive behavior is triggered when peer $p$ receives a gossip message containing Mapping-content 
and view entries from some peer $p''$. Peer $p$ answers by sending back a gossip message with its own 
Mapping-content and view information, and updates its local view  with via merge() and 
select\_recent(), and finally updates the local Mapping-content with respect to the new view as 
described previously.


We omit the pseudocode of the Gossiping mapping entries Algorithm for lack of space.

\section{Experimental Evaluation}~\label{sec:expe}
We first describe in Section~\ref{sec:setup} the system setup.
In Section~\ref{sec:quali}, we assess the quality and efficiency of our rewriting technique, also with 
respect to traditional query reformulation approaches.
We then focus on the scalability of our algorithms and their robustness with 
respect to network churn in Section~\ref{sec:scala}.

\subsection{Experimental Setup}~\label{sec:setup}
\noindent {\bf Dataset and mapping generation.}
We have conducted our evaluation in PeerSim~\cite{peersim}, an 
open source simulator for P2P protocols. 
In order to tweak our system at best, we implemented a pseudo-randomized generator of relational schemas. Indeed, none of the available relational 
datasets could provide us enough heterogeneity to distribute on a large 
number of nodes in the network. 
Thanks to this generator, no peer's schema is identical to any other and, as 
a consequence, mappings are all distinct. Moreover, every peer has at least one 
acquaintance, connected to it via a mapping. This ensures that there are not
semantically disconnected peers in the PDMS.

The generator leverages a dictionary of about $40$ names, ranging from table names to 
attribute names.  
We have designed a total of $10$ scenarios (outlined in Table~\ref{tab:exp}), by varying the number of peers in the PDMS, 
the number of mappings and the number of acquaintances, the latter ranging from a minimum of $1$
to a $21$, which is compatible with the diversity of the randomized schemas.    

\begin{table}[t!]
\centering
\scalebox{0.7}{
\begin{tabular}{|c|c|c|c|c|}
\hline
Scen. & \# of Peers  & \# of Mappings & \#Min Acq. & \#Max Acq.\\
\hline
\hline
1 & 500  &	2767  &	 3  &	12 \\
2 & 1000 &	6202  &	 2  &	14 \\
3 & 1500 &	9941  &  2  &	15 \\
4 & 2000 &	13814 &  2  &	16 \\
5 & 2500 &	17893 &  2  &	17 \\
6 & 3000 &	22037 &  1  &	18 \\
7 & 3500 &	26394 &	 1  &	18 \\
8 & 4000 &	30696 &	 1  &	20 \\
9 & 4500 &	34941 &	 1  &	21 \\
10 & 5000 &	39261 &	 1  &	21 \\ 
\hline
\end{tabular}
}
\vspace{-3mm}
\caption{Heterogeneous scenarios used for experiments.} \label{tab:exp}
\vspace{-5mm}
\end{table}



In the above scenarios, the number of mappings from a peer to each 
of its acquaintances ranges from $1$ to $6$, whereas each mapping has at most 
$3$ atoms in the body/head. Moreover, each peer's schema has been 
randomly generated to contain at most $6$ tables with at most $3$ attributes each.
The queries used in the experiments have been randomly generated to match the atoms in the body/head of 
the mappings, thus may contain in turn from $1$ to $3$ atoms.
Finally, the FOAF files are initially empty in all experiments, and are 
incrementally filled, as soon as query reformulation starts.

\noindent {\bf Qualitative measures and protocols for comparison.}
In each of the scenario depicted in Table~\ref{tab:exp}, one or more queries 
are formulated on initiating peers and they fire a certain number of 
\emph{relevant rewritings}, which represent all the rewritings for 
which the AF-IMF measure is greater than $0$. 
To evaluate the quality of the top-k mappings, we run our query 
reformulation algorithms in a centralized implementation of our protocol, and take the 
returned results for each query as relevant rewritings.
We have measured the \emph{recall}, which is computed as follows:

\vspace{-2mm}
\removesmallspace
\[
\begin{array}{l}

    \mathrm{Recall_{AF-IMF}} = \frac{Number of Retrieved Relevant Rewritings}{Total Number of Relevant Rewritings}\\

\end{array}
\removesmallspace
\]
 
Moreover, we have measured the time (in ms) taken to retrieve such relevant rewritings.  

In order to gauge the effectiveness of our techniques and also to 
provide a yardstick for comparison, we have implemented the following 
protocols, that have been used throughout the experimental assessment:

{\em Full} The query gets translated against the relevant (using AF-IMF) rewriting sequences, 
by exploiting LSV, gossiping and FOAF links.

{\em Full-} The query gets translated against the relevant (using AF-IMF) rewriting sequences, 
by exploiting LSV, gossiping (i.e. the protocol Full without FOAF links).

{\em Baseline$\#$} The query gets translated against the relevant (using AF only) rewriting sequences. 

{\em Baseline+} The original query gets translated against the mappings found in the traversal, 
and all its rewritings (relevant and non relevant) get propagated.

{\em Baseline} The original query gets translated against the mappings found in the traversal, 
and gets propagated as it is. 

With Baseline and Baseline+, we have reimplemented the propagation strategy of existing approaches~\cite{Halevy05,Bonifati2010}, adopting, however, the bidirectional translation semantics of our system. 

\noindent {\bf Initial System Setup.}
We have executed an initial set of experiments, aiming at determining the gossip thresholds
$VGossip$ and $LGossip$. The former indicated the size of the Mapping-content table in the LSV, while the latter 
allows to control the size of a gossip message within each gossip cycle. 
 Both parameters directly impact the effectiveness of the gossip protocol, since 
they indicate of what size an LSV and its buffer should be to harvest the highest number of relevant content in the network. 

From the experiments, that we omit for conciseness, we observe that a $VGossip$
size of $500$ entries is a good trade-off between number relevant rewritings 
retrieved and time, while varying the gossip cycles from $1$ to $10$.
We also observe that, if we keep $LGossip$ of the same size as $VGossip$ (in other words, we 
disseminate the entire LSV in gossip messages) or smaller, the results in terms of 
rewritings are not affected much. Therefore, we opted for a value of $LGossip = 100$
throughout the analysis.      

Moreover, as the ranking function to use in the 
TranslateQuery algorithm, we have adopted the harmonic mean, which overcomes by $0.5\%$ 
the other ranking functions (averaged on 
$10$ gossip cycles).
 
Finally, we have also empirically determined the maximum number of relevant requests $REQS$.
We observed in a batch of initial experiments that the number of rewritings is 
affected by a value of $REQS$ greater than $0$ only during the initial gossip cycles, 
whereas $REQS=0$ becomes the most preferable choice, 
when the number of gossip cycles increases. 
 From these experiments, we could infer that $REQS$ should be used as a dynamic threshold, and 
should have values slightly greater than $0$ when gossiping starts and 
drop to $0$ as long as gossip cycles reach $4$.

Also, we have set the threshold $\alpha$ of the number of query hops to \emph{unbounded}, to 
be able to observe the behavior of our algorithms in the most general case. 
Our prototype has been implemented in Java and all experiments have been performed on a 2.7 Ghz Intel Corei5 machine 
with 4GB RAM, running Windows 7 and JDK 6.
For all experiments (unless otherwise specified), we have used a PDMS of 
$2000$ peers with a configuration as in scenario $4$ of Table~\ref{tab:exp}.

\subsection{Qualitative Evaluation}~\label{sec:quali}
\begin{figure*}[t!]
\begin{center}
\begin{tabular}{cc}
\includegraphics[width=0.47\textwidth]{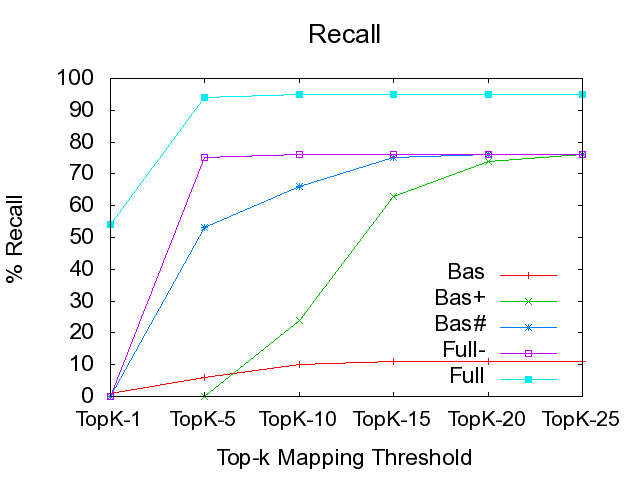}
&
\includegraphics[width=0.47\textwidth]{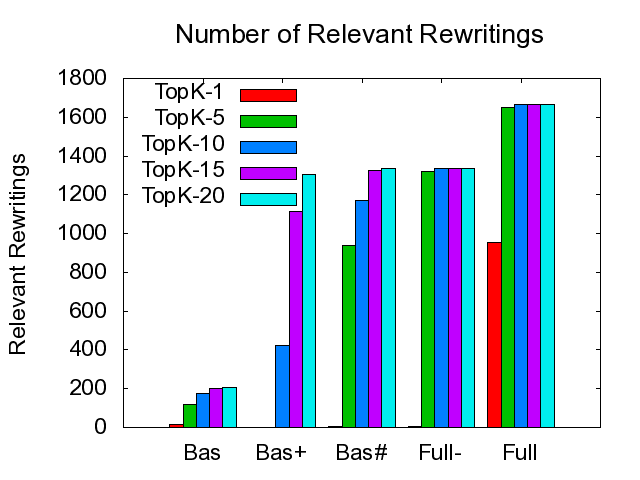} \\

\includegraphics[width=0.47\textwidth]{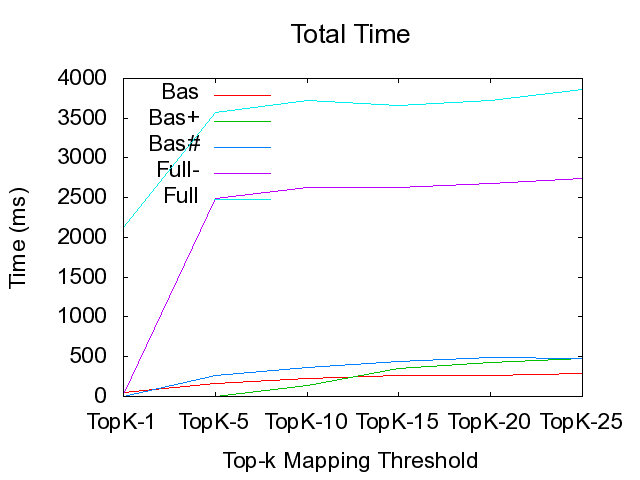}
&
\includegraphics[width=0.47\textwidth]{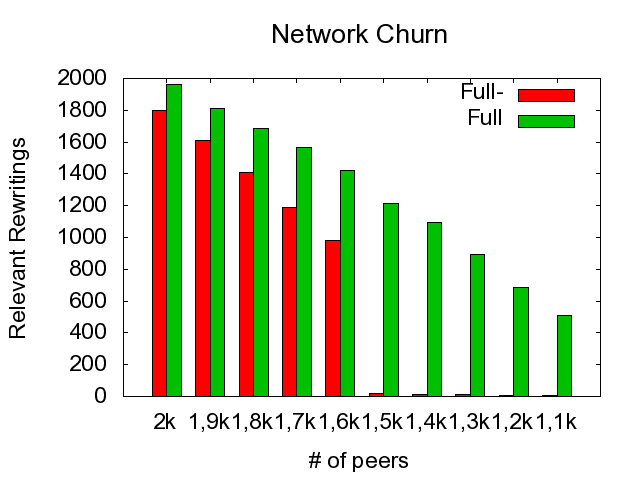}
\\
\end{tabular}
\caption{\small \label{fig:m1} 
(a) Recall, (b) $\#$ of Relevant Rewritings, (c) Total Time (ms) and (d) Network Churn from $2000$ peers down to $1100$ peers; $10$ queries.}
\end{center}
\end{figure*}

\noindent {\bf Recall and Comparison with previous approaches.}
As described above, we have measured the recall of our approach and compared it 
with the protocols \emph{Baseline} and \emph{Baseline+}. 
From Figure~\ref{fig:m1} (a), we can observe that our protocol Full has the greatest 
recall along all the values of top-k mappings, if compared with all other 
protocols. In particular, the contribution of FOAF links to 
the recall is noticeable, since such links enable to connect network areas which would be otherwise unexplored in the query translation process, 
as shown by the trend of Full and Full-.
Baseline, Baseline+ and Baseline$\#$ have a lower recall, as they do not 
exploit the relevance measure AF-IMF, thus the mappings that they exploit 
during query translation are in most cases not relevant. 
As long as more mappings are traversed, their recall improves, until Baseline$\#$
reaches the same recall of Full-, while it never reaches $100\%$ recall.
The latter is only achieved by the Full protocol, by exploiting the FOAF links.
Interestingly, this experiment showed the effectiveness of AF-IMF,
LSV and gossiping (from Baseline up to Full-) and the utility of FOAF links (from Full- to Full). 
In particular, it can be observed that adopting a local metric for evaluating mappings (like the AF metric of the Baseline$\#$ protocol) works better than using no metric at all (Baseline and Baseline+). However, it performs worse than using the AF-IMF global metric (like in Full- and Full), which has the most desirable behavior amongst all scenarios.

Similarly, in terms of the number of relevant rewritings, as shown in Figure~\ref{fig:m1} (b),
the Full protocol is the one that can harvest the highest number at any value of the top-k mappings.
Finally, we have quantified the cost incurred by the Full and Full- protocols 
with respect to the Baseline protocols. The results are reported in Figure~\ref{fig:m1} (c), 
which reports the time averaged on $10$ queries. We can observe that the times undertake 
a certain increase, due mainly to the gossiping active and passive threads, and to the 
computation of relevance for Full- and, additionally, to the FOAF linking for Full. 
However, these times are still reasonable as the latter protocols allow a 
significant increase of the recall (as shown in Figure~\ref{fig:m1} (a)).

\noindent {\bf Precision of distributed IMF.}
Next, we conducted another experiment to gauge the effectiveness of our 
distributed technique to compute AF-IMF.
We have defined the precision of distributed IMF as follows:
   
\removesmallspace
\[
\begin{array}{l}

    \mathrm{Precision_{IMF}} = \frac{Computed IMF Value}{Expected IMF Value}\\

\end{array}
\removesmallspace
\]
\vspace{1mm}

We recall that IMF only depends on the query and not on the mappings, whereas 
AF depends on the mappings. 
There are no false positives in the query reformulation 
process, therefore the precision of AF cannot be determined.
For such a reason, the precision we have measured is defined on IMF. 

By measuring such precision while varying the gossip cycles, we 
indirectly measure the effectiveness of the LSV. 
We can observe that in about $3$ gossip cycles,
the number of inquiries converges to $REQS=0$, meaning that the LSV has fetched enough relevant tuples from the outside and is self-contained. 
The \emph{backward} precision has a similar trend, and is omitted for lack of space. 

\begin{figure*}[t!]
\begin{center}
\begin{tabular}{cc}
\includegraphics[width=0.47\textwidth]{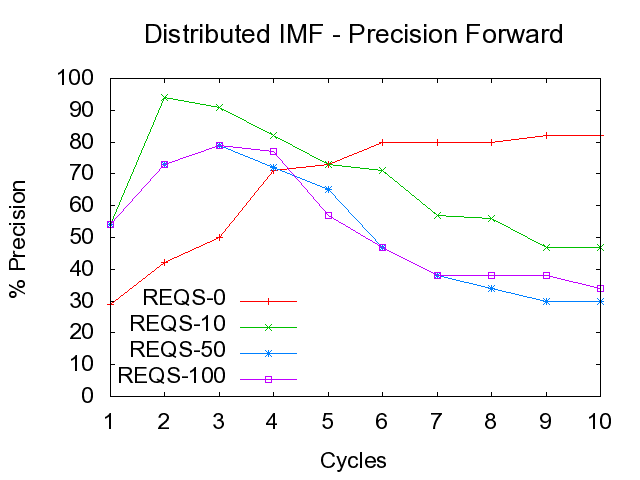}
&
\includegraphics[width=0.47\textwidth]{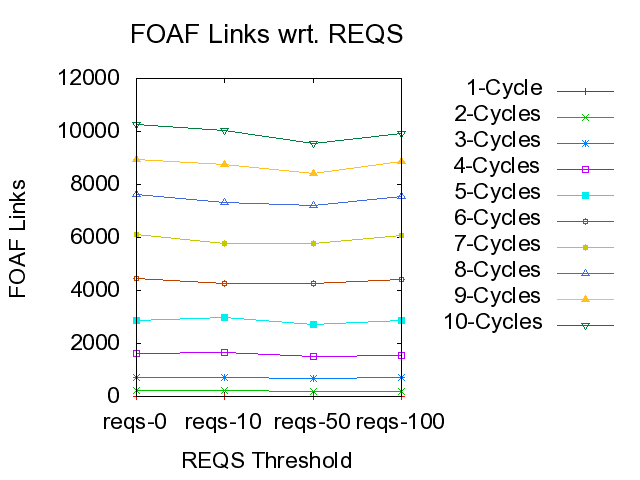} \\

\includegraphics[width=0.47\textwidth]{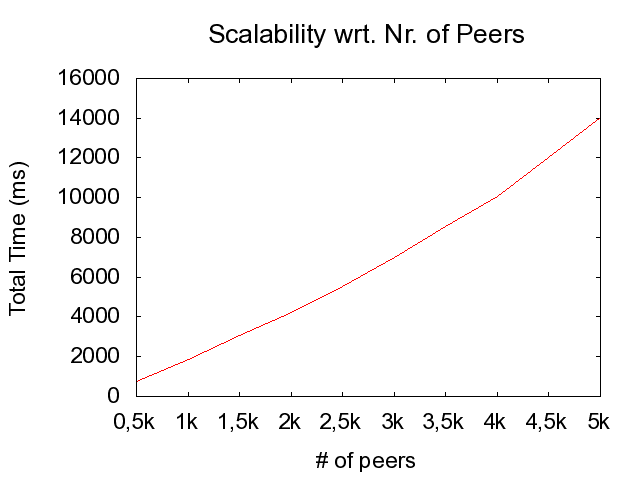}
&
\includegraphics[width=0.47\textwidth]{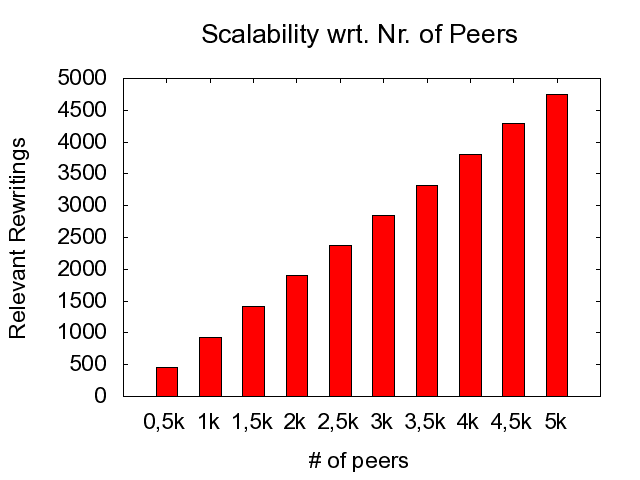}
\\
\end{tabular}
\caption{\small \label{fig:m2} 
(a) Distributed IMF Relevance Forward Precision, (b) Impact of $REQS$ on FOAF Links, (c) Scalability wrt. $\#$ of Peers (Time) and (d) Scalability wrt. $\#$ of Peers (Nr. of Rewritings); $10$ queries. }
\end{center}
\vspace{-9mm}
\end{figure*}

\noindent {\bf Effectiveness of FOAF links.}
Figure~\ref{fig:m2} (b) shows another experiment we have conducted to gauge the increase of the number of FOAF links as the number of gossip cycles grows. Such increase is not affected much by the threshold $REQS$, thus confirming that the converging value of $REQS = 0$ already conveys enough FOAF links.

\subsection{Scalability and Churn}~\label{sec:scala}
We now assess the robustness of our techniques 
in large-scale PDMS, by varying the number of peers (spanning 
all scenarios in Table~\ref{tab:exp}).
In Figure~\ref{fig:m2} (c) and (d), both the average time and $\#$
of relevant rewritings have a linear growth as the number of peers increases. 
This confirms that our techniques are scalable. 

In the next experiment, we have simulated the network churn, by starting 
from an initial configuration of $2000$ peers, and forcing $100$ peers at a time to 
leave the network. The aim of this experiment was twofold, to measure the robustness of 
our PDMS to churn, and to show the utility of FOAF links in a situation in which acquantainces of 
the peers (along with their mappings) quit the network. 

Figure~\ref{fig:m1} (d) shows that the Full- protocol (without FOAF) gets a few 
relevant rewritings after a cutoff point, i.e. when the $\#$ of peers drops to $1500$; 
indeed, the useful acquaintances have left, 
and no FOAF link can be exploited to get new rewritings. Conversely, the Full 
protocol scales gracefully as the $\#$ of peers decreases and exhibit 
a linearly decreasing number of rewritings, thus showing the utility of
FOAF links. 



\section{Related Work}~\label{sec:rel}
There has been a great deal of work on data management in 
P2P databases on issues ranging from schema mediation~\cite{Halevy05}
to mapping data values~\cite{miller03}, query processing~\cite{Koloniari04}
and query translation~\cite{Halevy05,Bonifati2010}.

Kementsietsidis et al.~\cite{miller03} describes a set of algorithms for 
exchanging data among peers, by only leveraging constraints on 
such exchange under the form of mapping tables, that comprise data values of the local peer and 
of external peers. Constraints on the content of peers under the form 
of logical rules are also studied in theoretical work on data integration~\cite{Lenzerini2002}. 

The only previous work that considered query reformulation in this context is~\cite{Halevy05,Bonifati2010}. 
In Piazza~\cite{Halevy05}, each peer is equipped with inclusion and equality mappings and 
a set of local storage 
descriptions.
Query answering is done by evaluating the containment of any arbitrary 
external conjunctive query against the mappings and the storage descriptions. 
However, no approximation of the local peer mappings with suitable 
data structures is adopted. Moreover,~\cite{Halevy05}
relies on a centralized index rather than on a distributed one.
 A schema mapping and query 
translation framework for XML databases is presented in both~\cite{Yu2004, Bonifati2010}, which disregards the problem of ranking mappings based on relevance, as we do in this paper. 

We focus on individual rewritings in this work and 
adopt the query rewriting semantics of~\cite{Bonifati2010,Fagin200589}.
Query rewriting with respect to a set of views is addressed in Minicon~\cite{potti01}, 
where views are joined to return the maximally contained rewritings for LAV
data integration. 

Data integration in the presence of a global mediated ontology, relational data sources and GAV mappings is also addressed in~\cite{Calvanese08}. 

Efficient XML query processing in P2P~\cite{Koloniari04}, 
leveraged multi-level Bloom-filters.
However, we are not focusing on query optimization for XML.

Finally, Kantere et al.~\cite{Kantere09} consider the problem of clustering peers 
based on their common interests in unstructured networks. 
Contrarily to our approach, 
they utilize metrics to compare a query and its rewritings, that are applied after the 
rewritings have been computed and not  
beforehand, as in our approach. 
 Moreover, our global AF-IMF metric is the first to take into account the entire collection of mappings in the network. 
The idea of quantifying the information transfer of individual schema mappings with local metrics is the subject of 
recent work~\cite{Arenas2010}. However, no global metrics in a social and distributed context are considered. 



Gossiping as a mean to enter diverse semantic domains is
used in~\cite{Aberer03}, where basically mappings between peers may not be correct 
or simply not be aligned with a given domain. Therefore, the paper shows how local 
mappings can be used to establish a global semantic agreement among the peers.

\section{Conclusions and Future Work}~\label{sec:concl}
To conclude, in this paper we have studied the problem of semantic query reformulation in 
social PDMS. We have presented a new notion of relevance of a query to a mapping and introduced a global metric for ordering the mappings considered in query rewriting. 
Future work is devoted to study the impact of query personalization, the combination with other quality metrics and the extension to unions of conjunctive queries (UCQs).


\bibliographystyle{plain}
\bibliography{report}

\end{document}